\documentclass[9pt,twocolumn,twoside]{opticajnl}
\journal{opticajournal} 

\setboolean{shortarticle}{false}

\usepackage{lineno}

\title{Beam angle-locked loop in laser interferometry}

\author[1,2,*]{Daikang Wei}
\author[1,2]{Christoph Bode}
\author[1,2]{Vitali M\"{u}ller}
\author[1,2]{Juan Jos\'e Esteban Delgado}
\author[1,2]{Gerhard Heinzel}

\affil[1]{Max-Planck-Institut f\"ur Gravitationsphysik (Albert-Einstein-Institut), Callinstraße 38, 30167 Hannover, Germany}
\affil[2]{Leibniz Universit\"at Hannover, Institut für Gravitationsphysik, Callinstraße 38, Hannover 30167, Germany}

\affil[*]{daikang.wei@aei.mpg.de}

\begin{abstract}
Maintaining precise beam coalignment is essential in laser interferometry, particularly for interspacecraft missions where angular misalignments introduce tilt-to-length coupling noise and risk complete link failure. This work presents a comprehensive theoretical and experimental framework for the beam angle-locked loop, which maintains coalignment between two interfering beams. Minute angular misalignments are detected via differential wavefront-sensing signals, which are used as the error signal for feedback control. A detailed linear control model incorporating an angle detector, a digital filter, a proportional-integral-double-integral servo, a digital-to-analog converter, and a beam-steering mechanism is developed. Using loop transfer functions, we establish a noise-propagation model that quantifies the contributions of individual components to both out-of-loop and in-loop angle errors. The analytical results are experimentally validated using a transponder-based interferometric link spanning two optical benches, with a hexapod simulating spacecraft attitude jitter. The measured transfer functions and coalignment performance agree well with theoretical predictions, yielding a pointing stability better than 10 µrad$/\sqrt{\text{Hz}}$ between 0.2 mHz and 1 Hz. Furthermore, a loop-optimization strategy is demonstrated to minimize angle misalignment within a target frequency band by tuning the servo gain, without requiring hardware modifications. This validated framework provides a reliable architecture for designing and optimizing active beam alignment in laser interferometry.
\end{abstract}

\setboolean{displaycopyright}{false} 

\begin{document}

    \maketitle

    \section{Introduction}
    Laser interferometry is a highly sensitive technique for detecting minute distance variations through corresponding changes in optical path length, facilitating precision measurements in experimental gravity and gravitational-wave astronomy. The most prominent ground-based laser interferometer is the Laser Interferometer Gravitational-Wave Observatory (LIGO), which achieved the first direct detection of gravitational waves, thereby opening the window to gravitational-wave astronomy \cite{abbott2016observation, ligo2015advanced}. In space-based missions, such as the Laser Interferometer Space Antenna (LISA), the Gravity Recovery and Climate Experiment Follow-On (GRACE-FO), GRACE-Continuity (GRACE-C), and the Next Generation Gravity Mission (NGGM), laser interferometers are employed to measure infinitesimal variations in the distance between spacecraft \cite{LISA, abich2019orbit,landerer2024towards, daras2024next}.
    
    Precise coalignment of the interfering beams is a critical requirement for laser interferometry. In ground-based power-recycled interferometers, precise beam coalignment ensures optimal mode matching across optical cavities and interferometer arms while minimizing the coupling of alignment and geometric fluctuations into the light power \cite{fritschel1998alignment,grote2004alignment}. In interspacecraft interferometry, angular misalignment between the receiving (RX) and transmitting (TX) beams introduces measurement noise, a critical noise source known as tilt-to-length (TTL) coupling \cite{hartig2023non,hartig2022geometric}. Misalignments near or above the beam divergence will lead to a complete loss of the laser interferometric link.
    
    To mitigate misalignment effects, active beam alignment systems are required in laser interferometry. A standard technique for detecting minute angular misalignments is differential wavefront sensing (DWS). By analyzing the phase differences between the segments of a quadrant photoreceiver (QPR), DWS provides a highly sensitive measure of the wavefront tilt between two interfering beams. This signal can be utilized as feedback in a control loop to steer the beam and maintain coalignment. This control loop, which maintains the alignment of one beam relative to another, is referred to as a beam angle-locked loop (BALL) in this work.

    The interferometric system, which includes a phasemeter for precise beam angle readout, constitutes the angle detector in the BALL. As the output of the angle detector,  the DWS signal is propagated through a control chain typically comprising a low-pass filter (LPF), a servo system, a digital-to-analog converter (DAC), and a beam-steering mechanism (BSM). By processing this signal, the BSM actively steers the beam to achieve angle locking relative to the reference beam.

    The DWS-based control loop was first developed for automatic alignment in ground-based gravitational wave detectors \cite{morrison_experimental_1994,morrison_automatic_1994}. This automatic alignment system has since been widely applied to ground-based interferometers \cite{fritschel1998alignment,heinzel_automatic_1999,grote2004alignment}. The BALL has been successfully implemented in the GRACE-FO laser ranging interferometer (LRI), where a fast steering mirror (FSM) is used to compensate for spacecraft attitude jitter, ensuring coalignment between the RX and TX beams \cite{sheard2012intersatellite,abich2019orbit}. A similar concept will be adopted in future interspacecraft interferometry missions, including GRACE-C, NGGM, and LISA. In the LISA mission, however, unlike the GRACE-FO LRI, the DWS signals are fed directly back to the spacecraft's drag-free and attitude control system (DFACS) rather than to an FSM. To optimize beam coalignment performance and support future laser interferometer missions, it is necessary to establish a theoretical framework for the BALL that provides a rigorous understanding of the transfer functions and noise contributions of the individual components, ranging from optical detection to the digital servo and mechanical actuator. This framework is directly applicable to the design of active beam alignment systems in laser interferometry.

    In this paper, a comprehensive theoretical and experimental analysis of the BALL for laser interferometry is presented. The DWS signal, which senses beam misalignment, forms the basis of angle detection in the BALL. A detailed linear model of the control loop is developed that incorporates all loop components. By integrating the loop transfer function, a noise model is established that quantifies the contributions of reference-beam jitter and individual BALL components to the total angle error, for both out-of-loop and in-loop configurations. To validate the model, we conducted an experiment using a transponder-based laser interferometric link spanning two optical benches. The measured transfer functions and beam coalignment performance agree well with theoretical predictions. Additionally, the developed model was used to optimize the BALL by tuning the servo gain to minimize angle misalignment while maintaining loop stability. 
    
    The paper is organized as follows. Section~\ref{ALL} outlines the theoretical framework of the BALL, encompassing the principles of angle detection via DWS, the linear loop model, and the associated noise analysis. The experimental work is presented in Sec.~\ref{Experiment}, which provides a comprehensive overview of the experimental setup, the characterization of transfer functions, noise and coalignment measurements, and loop optimization strategies. The paper concludes with a discussion and final remarks in Sec.~\ref{dis_con}.

    \section{Angle-locked loop}\label{ALL}
    \subsection{Angle detection}\label{angle_detection}

    \begin{figure}[!htp]
		\centering
		\includegraphics[width=0.48\textwidth]{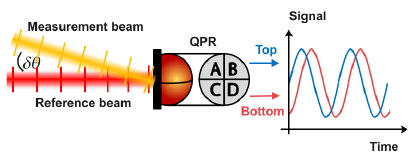}
		\caption{Schematic of the differential wavefront sensing technique. An angular tilt of
        $\delta\theta$ is applied to the measurement beam (yellow) relative to the reference beam (red). The beams interfere on the surface of a quadrant photoreceiver (QPR), generating heterodyne signals at each segment. The pivot point of the tilt is centered on the QPR, and there is no displacement between the beam centers. The angular tilt shown here is exaggerated for visualization. The plot on the right displays the average phase of segments A and B (blue curve) and segments C and D (red curve).}
		\label{DWS_explain}
    \end{figure}

    DWS is a standard technique in laser interferometry for quantifying minute angular misalignments between two beams \cite{morrison_experimental_1994,morrison_automatic_1994,armano2016sub,heinzel2020tracking}. As illustrated in Fig.~\ref{DWS_explain}, two laser beams with a relative tilt impinge on a QPR. A vertical angular misalignment between the wavefronts induces a phase shift between the heterodyne signals measured from the top (A and B) and bottom (C and D) regions. Similarly, a horizontal angular misalignment produces a phase difference between the left (A and C) and right (B and D) regions. Based on the measured phases $\varphi_{A}$,  $\varphi_{B}$,  $\varphi_{C}$, and  $\varphi_{D}$, the horizontal and vertical DWS signals are calculated by 
    \begin{eqnarray}
        \label{DWS_h}
        \mathrm{DWS_h}  & = & \frac{\varphi_{A}+\varphi_{C}-\varphi_{B}-\varphi_{D}}{2}, \\
        \label{DWS_v}
        \mathrm{DWS_v} & = & \frac{\varphi_{A}+\varphi_{B}-\varphi_{C}-\varphi_{D}}{2}.
    \end{eqnarray}

    In practical interferometry systems, imaging optics are typically employed to suppress beam walks and TTL coupling by imaging the tilt pivot onto the center of the QPR \cite{schuster2016experimental,trobs2018reducing,chwalla2020optical}. Here, we consider a scenario in which the tilt pivot is located at the center of the QPR and the beam centers are coaligned, which differs from the more general cases discussed in \cite{hechenblaikner2010measurement, wanner2012methods,pizzella2025mathematical}. For a small misalignment angle $\delta\theta_{\text{h/v}}$, the DWS signal can be approximated by
    \begin{equation}
        \label{DWS_coupling}
        \mathrm{DWS_{h/v}}  \approx \kappa \cdot \delta\theta_{\mathrm{h/v}}, \\
    \end{equation}
   where $\kappa$ is the linear angle-coupling factor derived from a first-order approximation \cite{hechenblaikner2010measurement, wanner2012methods,pizzella2025mathematical}. To analyze this coupling factor, we define the effective spot radius $w_{\mathrm{eff}}$, the relative wavefront curvature $R_{\mathrm{rel}}$, and a quantity $\rho$ to describe their relationship:
    \begin{eqnarray}
        \label{effective_spot}
        \frac{2}{w^{2}_{\text{eff}}} &=\frac{1}{w^{2}_{\text{r}}} + \frac{1}{w^{2}_{\text{m}}}, \\
        \label{relative_wavefront_r}
        \frac{1}{R_{\mathrm{rel}}} &= \frac{1}{R_{\mathrm{r}}} - \frac{1}{R_{\mathrm{m}}},\\
        \label{quantity_sigma}
        &\rho =\frac{k \cdot w^{2}_{\mathrm{eff}}}{4R_{\mathrm{rel}}},
    \end{eqnarray}
    where $w_{\text{r}}$ and $w_{\text{m}}$ denote the spot radii of the reference and measurement beams on the QPR, respectively; $R_{\text{r}}$ and $R_{\text{m}}$ denote their respective wavefront curvatures; and $k$ denotes the angular wavenumber of the laser beam. For the case of an infinite QPR (where the QPR radius $r_{\text{QPR}} \gg  w_{\mathrm{eff}}$),  the angle-coupling factor can be expressed as 
    \begin{equation}
        \label{infinite_qpd}
        \kappa  = \sqrt{\frac{2}{\pi}} \cdot k \cdot w_{\mathrm{eff}} \cdot \sqrt{\frac{1+\sqrt{1+\rho^2}}{2\left( 1+ \rho^2\right)}},
    \end{equation}
    which is a function of $k$, $w_{\mathrm{eff}}$, and $\rho$\cite{pizzella2025mathematical}. For the more general case of a finite QPR, the linear relation still holds, and the angle-coupling factor is given by
    \begin{eqnarray}
        \label{finite_qpd}
        \kappa = &k \cdot w_{\mathrm{eff}} \cdot \left( F_{0}\left(\eta \right) + F_{2}  \left(\eta \right) \cdot \rho^2
        \right),\\
        \label{beta_qpr}
        &\eta = \frac{\sqrt{2} r_{\text{QPR}}}{w_{\mathrm{eff}}},
    \end{eqnarray}
    which represents a second-order expansion of $\rho$. Here, $\eta$ describes the relative magnitude of the QPR active area to the effective beam size on the QPR. The detailed funitions of $F_0\left(\eta\right)$ and $F_2\left(\eta\right)$ are provided in Appendx \ref{app_angle_detection} (\eqref{F_0_beta} and \eqref{F_2_beta}). Further process regarding the mathematical derivation of the DWS angle-coupling factor is reported in Ref.~\cite{pizzella2025mathematical}. The DWS angle-coupling factor depends on both the parameters of the interference beams and the QPR area. Generally, $\kappa$ is on the order $10^3$ to $10^4$, making DWS signals highly sensitive for the detection of tiny angular misalignments.

    \subsection{Linear model}\label{linear_model}
    \begin{figure}
		\centering
		\includegraphics[width=0.482\textwidth]{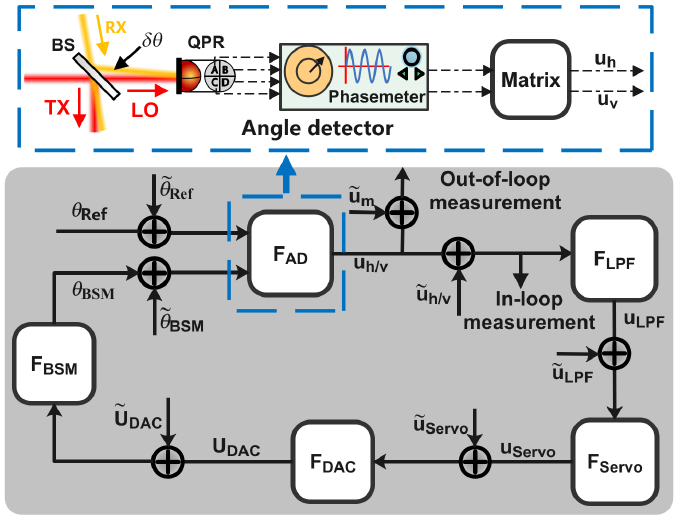}
		\caption{Block diagram of the linear beam angle-locked loop model. The principle of the angle detector is illustrated in the blue-dashed box. $u$, $U$, and $\theta$ denote the digital signal, analog signal, and beam angle, respectively, while the corresponding noise terms are represented by $\widetilde{u}$, $\widetilde{U}$, and $\widetilde{\theta}$. $\theta_{\text{Ref}}$ and $\theta_{\text{BSM}}$ denote the incidence angles of the reference and steering beams on the QPR, respectively. This BALL actively maintains the coalignment between the RX and LO beams, as well as between the RX and TX beams. In the first type of beam-steering setup, with the entire optical bench rotated, the $\theta_{\text{Ref}}$ corresponds to the LO or TX beam angle, while the $\theta_{\text{BSM}}$ corresponds to the RX beam angle. In the second setup with an FSM beam steering, these roles are reversed: the RX beam becomes the reference beam, while the LO and TX beams serve as steering beams. In-loop and out-of-loop measurements can be implemented using balanced detection with two QPRs and a beam splitter. The matrix in the angle detector corrects for QPR missalignment effects. BS, beam splitter; RX, receiving beam; TX, transmitting beam; LO, local beam; QPR, quadrant photoreceiver; AD, angle detector; LPF, low-pass filter; DAC, digital-to-analog converter; Ref, reference; BSM, beam-steering mechanism.}
		\label{angle_ll}
	\end{figure}
    
    As explained in Section \ref{angle_detection}, the DWS signal provides an angular measure of the misalignment between two beams. This signal can be used as a servo signal to steer the beam in a closed-loop operation. This constitutes the BALL concept, which actively maintains the coalignment of the two beams. This section discusses the linear model of the BALL. To achieve full beam steering control, two independent loops based on the $\mathrm{DWS_h}$ and $\mathrm{DWS_v}$ signals must be implemented. While the beam angle discussed here is referenced to the incident angle at the QPR, it can alternatively be referenced to the conjugate image plane of the QPR in applications employing imaging systems.
    
    The linear loop model presented here is adapted from the phase-locked loop (PLL) model \cite{gardner2005phaselock,gerberding2013phasemeter}. Fig.~\ref{angle_ll} presents the block diagram of the linear BALL and illustrates the principle of the angle detector. The loop comprises both analog and digital control components. The digital control is implemented using fixed-point arithmetic in a field-programmable gate array (FPGA). The subsequent analysis is based on the Laplace domain for analog components and the $z$ domain for digital components. The connection between the two domains is established via the transform $z = e^{s/f_s}$, where $s$ represents the complex frequency variable and $f_s$ represents the sampling frequency.

    As shown in the blue-dashed region of Fig.~\ref{angle_ll}, the RX and local oscillator (LO) beams are combined at a beam splitter (BS) to generate a heterodyne interference signal. This signal is detected by a QPR as a relative misalignment angle $\delta\theta$. The signals from each segment are processed by a phasemeter to obtain DWS signals. Since the unfolded TX beam is aligned with the LO beam at the QPR, the measured DWS signals reflect the misalignment between the RX and LO beams, as well as between the RX and TX beams. Accordingly, the BALL, based on the DWS signals, actively maintains coalignment for both beam pairs. Following the phasemeter, a matrix corrects DWS cross-coupling between the horizontal and vertical axes induced by QPR misalignment. This matrix is determined experimentally via a dedicated rotation calibration \cite{wei2026experimental}. Further details regarding this matrix are provided in Appendix \ref{app_DWS matrix}. Assuming the DWS signal, which spans $[-\pi, \pi]$, is digitized to an $X$-bit resolution, the linear transfer function of the angle detector can be expressed using \eqref{DWS_coupling} as
    \begin{equation}
        F_{\text{AD}} = \frac{\mathrm{u_{h/v}}}{\delta \theta} = \frac{\kappa 2^X}{2\pi},
    \label{F_AD}
    \end{equation}
	which has a unit of [1/rad]. $\mathrm{u_{h/v}}$ represents the digital output of the angle detector.

    Following the angle detector, an LPF suppresses unwanted high-frequency components, particularly high-frequency ripple and spurs. In the case of piezo actuators used for beam steering, this LPF prevents them from operating near their mechanical resonant frequency, thereby avoiding potential damage. The filter design and implementation depend on the specific loop configuration and must be adapted accordingly. Using the LPF output $\mathrm{u_{LPF}}$, the dimensionless gain of the LPF is given by
    \begin{equation}
    F_{\text{LPF}}\left(z\right) = \frac{\mathrm{u_{LPF}}}{\mathrm{u_{h/v}}}.
    \label{F_LPF}
    \end{equation}

    The servo system consists of two stages: a constant-gain scaling and a proportional–integral–double-integral (PII) controller. The gain scaling is implemented via signal bit-width expansion, which prevents overflow in the accumulators and preserves numerical precision in the subsequent controller. Instead of a general proportional-integral controller, a PII controller is selected to further improve low-frequency gain. Hence, the transfer function of the servo system is described by
    \begin{equation}
    \begin{aligned}    
        F_{\text{Servo}}\left(z\right) &= \frac{\mathrm{u_{Servo}}}{\mathrm{u_{LPF}}} \\
        &= 2^{-Y} \cdot\left( g_p+g_i\frac{z^{-1}}{1-z^{-1}}+g_{ii}\frac{z^{-2}}{\left(1-z^{-1}\right)^2}\right),
    \end{aligned}
    \label{F_servo}
    \end{equation}
    where $\mathrm{u_{Servo}}$ denotes the digital output of the servo system, $Y$ denotes the number of bits added during the gain scaling stage, and $g_p$, $g_i$, and $g_{ii}$ denote the gain factors of the proportional, integral, and double integral controllers, respectively. Since both $u_{\text{Servo}}$ and $u_{\text{LPF}}$ are digital signals, $F_{\text{Servo}}$ is a dimensionless function. Note that the accumulators used in \eqref{F_servo} introduce a one-clock-cycle input delay.

    A DAC converts the digital signal into an analog signal that drives the actuator later. The DAC's transfer function depends on its specification. In general, its transfer function is approximated by a constant for frequencies well below its bandwidth. Using the DAC input ($\mathrm{u_{Servo}}$) and output ($\mathrm{U_{DAC}}$), the transfer function of the DAC is provided by
    \begin{equation}
        F_{\text{DAC}}\left(s\right) = \frac{\mathrm{U_{DAC}}}{\mathrm{u_{Servo}}},
    \label{F_DAC}
    \end{equation}
    where the value represents the conversion of a digital code to a voltage, resulting in a unit of [V] for $F_{\text{DAC}}\left(s\right)$.

    The actuator in the loop is the BSM, which is driven by the DAC output. There are two types of BSMs. The first involves rotating the entire optical bench to suppress the RX beam jitter while keeping the LO and TX beams static. The second mechanism uses an FSM installed on the optical bench. The FSM rotates the LO and TX beams to track the RX beam, maintaining coalignment between them. A typical example of the first type is the LISA mission \cite{LISA}, which employs a DFACS to maintain beam pointing. An example of the second type is the GRACE-FO LRI \cite{sheard2012intersatellite}, where a DWS-based beam steering loop maintains the coalignment between the RX and TX beams. 
    The two types of BSMs can also be combined in a nested scheme \cite{wei2026nestedactivepointingcontrol}, with the DFACS providing slow optical-bench rotation and the FSM performing faster corrections near its central operating position.

    As illustrated in Fig.~\ref{angle_ll}, $\theta_{\text{Ref}}$ and $\theta_{\text{BSM}}$ denote the incident angles at the QPR for the reference and steering beams, respectively. In the first configuration, the reference beam corresponds to the LO or TX beam, while the steering beam corresponds to the RX beam. In the second configuration, these assignments are reversed. The transfer function of the BSM is written as
    \begin{equation}
    F_{\text{BSM}}\left(s\right) = \frac{\theta_{\text{BSM}}}{\mathrm{U_{DAC}}} =m \cdot K_{\text{BSM}}\cdot A(s),
    \label{F_BSM}
    \end{equation}
    where $m$ denotes the angular magnification of the imaging system, $K_{\text{BSM}}$ denotes the sensitivity coefficient between the BSM and the beam angular deflection, and $A(s)$ denotes the frequency response of the BSM to an input voltage signal. The unit of $F_{\text{BSM}}$ is [rad/V]. An imaging system is typically used in interferometry to suppress beam walk and TTL coupling; its magnification, denoted $m$, is considered here. The sensitivity coefficient $K_{\text{BSM}}$ depends on the type of BSM. For the first type, where the entire optical bench is rotated, $K_{\text{BSM}} = 1$. For the second type, based on an FSM, $K_{\text{BSM}}$ is determined by the incident angle $\theta_i$ and the beam-steering direction. $K_{\text{BSM}}$ is given by the sensitivity matrix, with a value of 2 for the horizontal direction (yaw) or $2\cos\theta_i$ for the vertical direction (pitch). Further discussion regarding the FSM $S_{\text{BSM}}$ is provided in Appendix \ref{app_fsm_bsm}, which details the mathematical derivation of the sensitivity matrix.
    
    Although not shown in Fig.~\ref{angle_ll}, the latency associated with signal processing and propagation must be included in the loop model. The corresponding transfer function is 
    \begin{equation}
    F_{\text{D}}\left(s\right) = e^{-s \tau},
    \label{F_D}
    \end{equation}
    where $\tau$ is the total loop delay. Combining \eqref{F_AD}, \eqref{F_LPF}, \eqref{F_servo}, \eqref{F_DAC}, \eqref{F_BSM}, and \eqref{F_D}, the open-loop transfer function of the BALL $G(s,z)$ is obtained as
    \begin{equation}
    \begin{aligned}
    &G\left(s,z\right) = F_{\text{AD}} \cdot F_{\text{LPF}}\left(z\right) \cdot F_{\text{servo}}\left(z\right) \cdot F_{\text{DAC}}\left(s\right) \cdot F_{\text{BSM}}\left(s\right) \cdot F_{\text{D}}\left(s\right) 
    \end{aligned}
    \label{G_ol_tf}
    \end{equation}
    Accordingly, the closed-loop transfer function $H\left(s,z\right)$ and the error transfer function $E\left(s,z\right)$ are derived using the control loop relationships: 
    \begin{eqnarray}
        H\left(s,z\right) = \frac{G\left(s,z\right)}{ 1+G\left(s,z\right)},
        \label{H_cl_tf}\\
        E\left(s,z\right) = \frac{1}{1+G\left(s,z\right)}.
        \label{E_tf}
    \end{eqnarray}

    \subsection{Noise analysis}\label{noise_model}
    
    As shown in Fig.~\ref{angle_ll}, various noise sources couple into the BALL at different stages, ultimately degrading the output beam pointing stability. In this section, we quantify these noise contributions using the linear model developed in Section \ref{linear_model}. For noise analysis, the relevant noise sources are classified according to their injection points as either sensing or actuation noise. Sensing noise enters through the error signal and is transferred to the output through the closed-loop transfer function $H(s,z)$, whereas actuation noise is suppressed by the error transfer function $E(s,z)$. 

    Assuming that all noise sources are statistically independent, each contribution is referred to an equivalent sensing or actuation noise and propagated to the output through the corresponding transfer function. All contributions ultimately manifest as equivalent angular noise in the output beam in the Laplace domain.
    
    The reference beam jitter, denoted by $\widetilde{\theta}_{\text{Ref}}$, represents angular fluctuations of the reference beam. Its contribution to the output beam angle is obtained by propagation through the closed-loop transfer function,
    \begin{equation}
       \widetilde{\theta}_{\text{Ref}}^{\text{out}}\left(s\right) =  H\left(s,z\right) \cdot \widetilde{\theta}_{\text{Ref}}.
       \label{noise_ref}
    \end{equation}

    The laser noise, QPR noise, and phasemeter noise contribute to the uncertainty of the angular readout, denoted by $\widetilde{u}_{\text{h/v}}$. In the linear model, this readout noise can be referred to the loop input as an equivalent sensing noise $\widetilde{u}_{\text{h/v}}/F_{\text{AD}}$. Its contribution to the output beam angle is therefore
    \begin{equation}
       \widetilde{\theta}_{\text{AD}}^{\text{out}}\left(s\right) =  \frac{H\left(s,z\right)}{F_{\text{AD}}} \cdot \widetilde{u}_{\text{h/v}}.
       \label{noise_AD}
    \end{equation}

    Digital filtering inevitably increases the bit length during the implementation. To conserve computational resources, truncation, rounding, or dithering may be applied. The noise introduced during this stage is defined by $\widetilde{u}_{\text{LPF}}$. Referring to the loop input, it corresponds to an equivalent sensing noise $\widetilde{u}_{\text{LPF}}/(F_{\text{AD}} \cdot F_{\text{LPF}}(z))$. The resulting output angular noise is
    \begin{equation}
       \widetilde{\theta}_{\text{LPF}}^{\text{out}}\left(s\right) =  \frac{H\left(s,z\right)}{F_{\text{AD}} \cdot F_{\text{LPF}}\left(z\right)}\cdot \widetilde{u}_{\text{LPF}}.
       \label{noise_LPF}
    \end{equation}

    The digital servo output drives the DAC. To match the DAC resolution, further quantization steps introduce noise $\widetilde{u}_{\text{Servo}}$. Referred to the loop input through the servo, LPF, and angle-detector responses, the corresponding equivalent sensing noise is $\widetilde{u}_{\text{Servo}}/(F_{\text{AD}} \cdot F_{\text{LPF}}(z) \cdot F_{\text{Servo}}(z))$. Its contribution to the output beam angle is
    \begin{equation}
       \widetilde{\theta}_{\text{Servo}}^{\text{out}}\left(s\right) =  \frac{H\left(s,z\right)}{F_{\text{AD}} \cdot F_{\text{LPF}}\left(z\right) \cdot F_{\text{Servo}}\left(z\right)} \cdot \widetilde{u}_{\text{Servo}}.
       \label{noise_Servo}
    \end{equation}

    Imperfections in the BSM introduce an angular uncertainty $\widetilde{\theta}_{\text{BSM}}$ for a given control input. Because this disturbance enters on the actuation side of the loop, it is suppressed by the error transfer function, yielding
    \begin{equation}
       \widetilde{\theta}_{\text{BSM}}^{\text{out}}\left(s\right) =  E\left(s,z\right) \cdot \widetilde{\theta}_{\text{BSM}}.
       \label{noise_BSM}
    \end{equation}
    Within the loop bandwidth, the steering beam tracks the reference beam, whereas high-frequency steering jitter remains uncorrected.

    Additional actuation noise originates from the DAC, including intrinsic electronic noise, power-supply ripple, reference-voltage fluctuations, and clock jitter. We denote the combined DAC output noise by $\widetilde{U}_{\text{DAC}}$. The BSM converts this voltage noise into angular fluctuations through its response $F_{\text{BSM}}(s)$, which are subsequently suppressed by the feedback loop. The resulting contribution to the output beam angle is
    \begin{equation}
       \widetilde{\theta}_{\text{DAC}}^{\text{out}}\left(s\right) =  E\left(s,z\right) \cdot  F_{\text{BSM}}\left(s\right) \cdot \widetilde{U}_{\text{DAC}}.
       \label{noise_DAC}
    \end{equation}

    Assuming statistical independence, the total output beam angle noise is obtained by linear superposition of all individual contributions:
    \begin{equation}
    \begin{aligned}
       \widetilde{\theta}_{\text{out}}\left(s\right) &= \widetilde{\theta}_{\text{Ref}}^{\text{out}}\left(s\right) + \widetilde{\theta}_{\text{AD}}^{\text{out}}\left(s\right) + \widetilde{\theta}_{\text{LPF}}^{\text{out}}\left(s\right) \\
       & + \widetilde{\theta}_{\text{Servo}}^{\text{out}}\left(s\right) +  \widetilde{\theta}_{\text{BSM}}^{\text{out}}\left(s\right)+ \widetilde{\theta}_{\text{DAC}}^{\text{out}}\left(s\right). 
    \end{aligned}
    \label{noise_output}
    \end{equation}
    The DWS signals are acquired using two independent detection paths: an in-loop measurement that provides feedback control, and an out-of-loop measurement that independently monitors the true beam coalignment. These two detections can be implemented with an optical BS and two QPRs. As illustrated in Fig.~\ref{angle_ll}, the out-of-loop angle error $\widetilde{\delta\theta}_{\text{out-loop}}(s)$ is given by
    \begin{equation}
    \begin{aligned}
       \widetilde{\delta\theta}_{\text{out-loop}}\left(s\right) &= \widetilde{\theta}_{\text{Ref}} - \widetilde{\theta}_{\text{out}}\left(s\right) + \frac{\widetilde{u}_{\text{m}}}{F_{\text{AD}}} \\
       & =E\left(s,z\right) \cdot (\widetilde{\theta}_{\text{Ref}}-\widetilde{\theta}_{\text{BSM}})- \frac{H\left(s,z\right)}{F_{\text{AD}}} \cdot \widetilde{u}_{\text{h/v}} \\
       &-  \frac{H\left(s,z\right)}{F_{\text{AD}} \cdot F_{\text{LPF}}\left(z\right)}\cdot \widetilde{u}_{\text{LPF}} \\
       &-\frac{H\left(s,z\right)}{F_{\text{AD}} \cdot F_{\text{LPF}}\left(z\right) \cdot F_{\text{Servo}}\left(z\right)} \cdot \widetilde{u}_{\text{Servo}} \\
       &-  E\left(s,z\right) \cdot  F_{\text{BSM}}\left(s\right) \cdot \widetilde{U}_{\text{DAC}} + \frac{\widetilde{u}_{\text{m}}}{F_{\text{AD}}},
    \end{aligned}
    \label{out_of_loop_error}
    \end{equation}
    where $\widetilde{u}_{\text{m}}$ denotes the readout noise of the independent measurement channel. In practice, loop errors are initially recorded as DWS signals and subsequently converted to angular deviations via \eqref{F_AD}. The error analysis presented here refers to the angular deviation rather than the DWS signals. In contrast, the in-loop error signal is directly fed back into the controller. Its expression is derived as
    \begin{equation}
    \begin{aligned}
       \widetilde{\delta\theta}_{\text{in-loop}}\left(s\right) &= \widetilde{\theta}_{\text{Ref}} - \widetilde{\theta}_{\text{out}}\left(s\right) + \frac{\widetilde{u}_{\text{h/v}}}{F_{\text{AD}}} \\
       & =E\left(s,z\right) \cdot (\widetilde{\theta}_{\text{Ref}}-\widetilde{\theta}_{\text{BSM}} + \frac{\widetilde{u}_{\text{h/v}}}{F_{\text{AD}}})\\
       &-  \frac{H\left(s,z\right)}{F_{\text{AD}} \cdot F_{\text{LPF}}\left(z\right)}\cdot \widetilde{u}_{\text{LPF}} \\
       &- \frac{H\left(s,z\right)}{F_{\text{AD}} \cdot F_{\text{LPF}}\left(z\right) \cdot F_{\text{Servo}}\left(z\right)} \cdot \widetilde{u}_{\text{Servo}} \\
       &-  E\left(s,z\right) \cdot  F_{\text{BSM}}\left(s\right) \cdot \widetilde{U}_{\text{DAC}}.
    \end{aligned}
    \label{in_loop_error}
    \end{equation}

    The power spectral density (PSD) of the output beam angle is obtained by multiplying the input noise PSDs by the squared magnitude of their respective transfer functions. Applying this principle to \eqref{out_of_loop_error} and \eqref{in_loop_error} yields the out-of-loop and in-loop angle misalignment PSDs, $S_{\delta\theta,\text{out}}(f)$ and $S_{\delta\theta,\text{in}}(f)$, respectively:
    \begin{equation}
    \begin{aligned}
       S_{\delta\theta,\text{out}}\left(f\right) & =|E\left(s,z\right)|^2 \cdot S_{\theta,\text{ref}}(f) + |E\left(s,z\right)|^2 \cdot S_{\theta,\text{BSM}}(f) \\
       &+ |\frac{H\left(s,z\right)}{F_{\text{AD}}}|^2 \cdot S_{u,\text{h/v}}(f) + |\frac{H\left(s,z\right)}{F_{\text{AD}} \cdot F_{\text{LPF}}\left(z\right)}|^2 \cdot S_{u,\text{LPF}}(f)\\
       & + |\frac{H\left(s,z\right)}{F_{\text{AD}} \cdot F_{\text{LPF}}\left(z\right) \cdot F_{\text{Servo}}\left(z\right)}|^2 \cdot S_{u,\text{Servo}}(f)\\
       &+ |E\left(s,z\right) \cdot  F_{\text{BSM}}\left(s\right)|^2 \cdot S_{U,\text{DAC}}(f) + |\frac{1}{F_{\text{AD}}}|^2 \cdot S_{u,\text{m}}(f),
    \end{aligned}
    \label{PSD_out_loop_error}
    \end{equation}
    \begin{equation}
    \begin{aligned}
       S_{\delta\theta,\text{in}}\left(f\right) & =|E\left(s,z\right)|^2 \cdot S_{\theta,\text{Ref}}(f) + |E\left(s,z\right)|^2 \cdot S_{\theta,\text{BSM}}(f) \\
       &+ |\frac{E\left(s,z\right)}{F_{\text{AD}}}|^2 \cdot S_{u,\text{h/v}}(f) + |\frac{H\left(s,z\right)}{F_{\text{AD}} \cdot F_{\text{LPF}}\left(z\right)}|^2 \cdot S_{u,\text{LPF}}(f)\\
       & + |\frac{H\left(s,z\right)}{F_{\text{AD}} \cdot F_{\text{LPF}}\left(z\right) \cdot F_{\text{Servo}}\left(z\right)}|^2 \cdot S_{u,\text{Servo}}(f) \\
       &+ |E\left(s,z\right) \cdot  F_{\text{BSM}}\left(s\right)|^2 \cdot S_{U,\text{DAC}}(f),
    \end{aligned}
    \label{PSD_in_loop_error}
    \end{equation}
    where $S_{\theta,\text{ref}}(f)$, $S_{\theta,\text{BSM}}(f)$, $S_{u,\text{h/v}}(f)$, $S_{u,\text{LPF}}(f)$, $S_{u,\text{Servo}}(f)$, $S_{U,\text{DAC}}(f)$, and $S_{u,\text{m}}(f)$ denote the PSDs of the reference beam jitter, BSM noise, angle detector noise, LPF noise, servo system noise, DAC output voltage noise, and measurement channel readout noise respectively. The complex frequency $s$ is related to the Fourier frequency $f$ via $s = 2\pi j f$.

    \section{Experiment}\label{Experiment}
    \subsection{Experimental setup}\label{experimental_setup}
\begin{figure*}[!htp]
		\centering
		\includegraphics[width=0.75\textwidth]{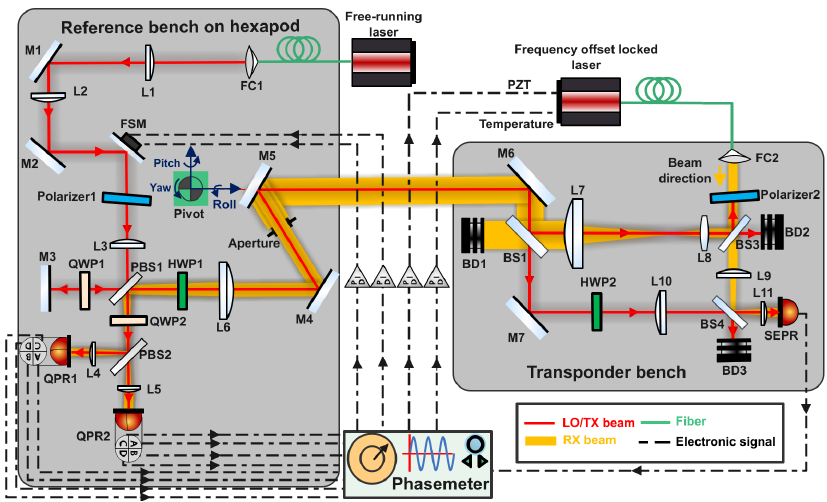}
		\caption{Schematic of the laser interferometric link spanning two optical benches. The reference bench is mounted on a hexapod, allowing manual introduction of beam jitter on the RX beam by rotating the entire bench. The coordinate frame attached to the reference bench defines the rotation axes. An FSM on the reference bench actively steers the TX and LO beams to track the RX beam from the transponder bench. Beam definitions (RX and TX) are referenced to the reference bench. The transponder bench is fixed and implements frequency-offset locking based on the TX beam from the reference bench. M, mirror; L, lens; FSM, fast steering mirror; QWP, quarter-wave plate; HWP, half-wave plate; SEPR, single-element photoreceiver; QPR, quadrant photoreceiver; FC, fiber collimator; BS, beam splitter; PBS, polarizing beam splitter; PZT, piezoelectric transducer; BD, beam dump.}
		\label{Experimental_setup}
    \end{figure*}

    This section presents the experimental study of the BALL. The experimental setup is based on an on-axis LRI concept \cite{wei2026experimental, yang2022axis, wei2026nestedactivepointingcontrol}, a candidate architecture for future gravity missions. Fig.~\ref{Experimental_setup} shows the experimental setup, which establishes a transponder-based laser interferometric link between the reference and transponder benches. The beam confined to the reference bench, which serves as the LO for heterodyne detection, is denoted by the LO beam. The beam propagating from the reference bench to the transponder bench is the TX beam, while the beam propagating from the transponder bench to the reference bench is the RX beam. 
    
    The reference bench is mounted on a hexapod (Newport, HXP100-MECA), which provides 3D rotation control with an accuracy of a few microradians \cite{NewportHXPManual}. Angular jitter in the RX beam ($\widetilde{\theta}_{\text{Ref}}$) can be manually introduced by rotating the bench about a dedicated pivot point. As illustrated in Fig.~\ref{Experimental_setup}, the rotation is defined by the coordinate frame attached to the reference bench, with the pivot point (the origin of the coordinate system) located at the mirror image of the aperture center. Utilizing the second type of BSM, an FSM is employed on the reference bench to steer the LO and TX beams, thereby maintaining beam-angle locking with the RX beam. In this experiment, the RX beam serves as the reference beam.
    
    A free-running laser provides a 5 mW beam at 1064 nm that enters the reference bench via a fiber. Lenses L1 and L2 image the beam waist onto an FSM (Physik Instrumente, S-325). A portion of the laser beam, with parallel polarization, passes through a polarizing beam splitter (PBS1), and its beam waist is further imaged onto QPR1 and QPR2. This portion of the laser serves as the LO beam. The remaining beam, constituting the TX beam, is directed to the transponder bench after its polarization is flipped by a combination of a mirror (M3) and a quarter-wave plate (QWP1). The beam waist of the TX beam is imaged onto an aperture. The power ratio between the LO and TX beams is controlled by rotating a linear polarizer (Polarizer1). The RX beam passes through the aperture and is imaged onto QPR1 and QPR2, producing heterodyne interference with the LO beam. Imaging systems are implemented to mitigate beam walks and TTL coupling caused by beam rotation, as well as to suppress diffraction effects at the aperture. With the implementation of imaging systems, the FSM, QPR1, QPR2, and the aperture form a set of conjugate image planes. The beam angle analyzed in this experiment is defined relative to the incident or exit angle at the aperture.

    The transponder scheme, originally developed for the LISA mission and demonstrated in the GRACE-FO mission \cite{jennrich2009lisa,abich2019orbit}, is implemented on the transponder bench. A portion of the laser beam on this bench interferes with the TX beam from the reference bench, generating heterodyne signals recorded by a single-element photoreceiver (SEPR). This signal is processed by a phasemeter to extract the heterodyne frequency. The measured frequency is fed back to control the laser's cavity length and crystal temperature, maintaining a stable frequency offset of 7.3 MHz between the two laser beams. The remaining laser beam, containing the majority of the power, is expanded by lenses L7 and L8 to a waist radius of approximately 21 mm. This expanded beam is sent to the reference bench as the RX beam. Upon reaching the reference bench, the RX beam clips at the aperture, and the clipped beam approximates a flat-field beam with a flat-top power intensity distribution.

    The phasemeter employs all-digital phase-locked loops (ADPLLs) to track the heterodyne signals, extracting the phase and frequency \cite{gerberding2013phasemeter}. By combining the measured phases from the different segments of the QPR, the DWS signals are obtained. In this experiment, angular jitter in the yaw and pitch directions was introduced simultaneously by controlling the hexapod rotations. To compensate for angular misalignment between the RX and TX beams, two independent BALLs based on the horizontal and vertical DWS signals measured from QPR1 were implemented. The measurement results from QPR1 are in-loop measurements, while those from QPR2 are out-of-loop measurements.

    \subsection{Transfer functions}\label{transfer_functions}    
    Two independent BALLs, based on the horizontal and vertical DWS signals, were implemented in the experimental setup (Fig.\ref{Experimental_setup}) and are referred to as the horizontal and vertical loops, respectively. Given the transfer functions of the individual components illustrated in Fig.\ref{angle_ll}, the loop transfer function can be derived using the linear model developed in Section \ref{ALL}. Alternatively, the loop transfer function can be characterized experimentally using a frequency-swept sine signal. This section compares the measured and modeled open-loop transfer functions.

    As indicated in \eqref{F_AD}, the transfer function of the angle detector is determined by the angle-coupling factor $\kappa$. This coupling factor was experimentally determined using the angle calibration method described in Ref.~\cite{wei2026experimental}. The hexapod executed angular scans with known angles, manually introducing angular misalignment in the RX beam. The coupling factor between the rotation angles and the DWS signals was determined by fitting the measured DWS angles to the rotation angles.

\begin{figure}[!htp]
		\centering
		\includegraphics[width=0.483\textwidth]{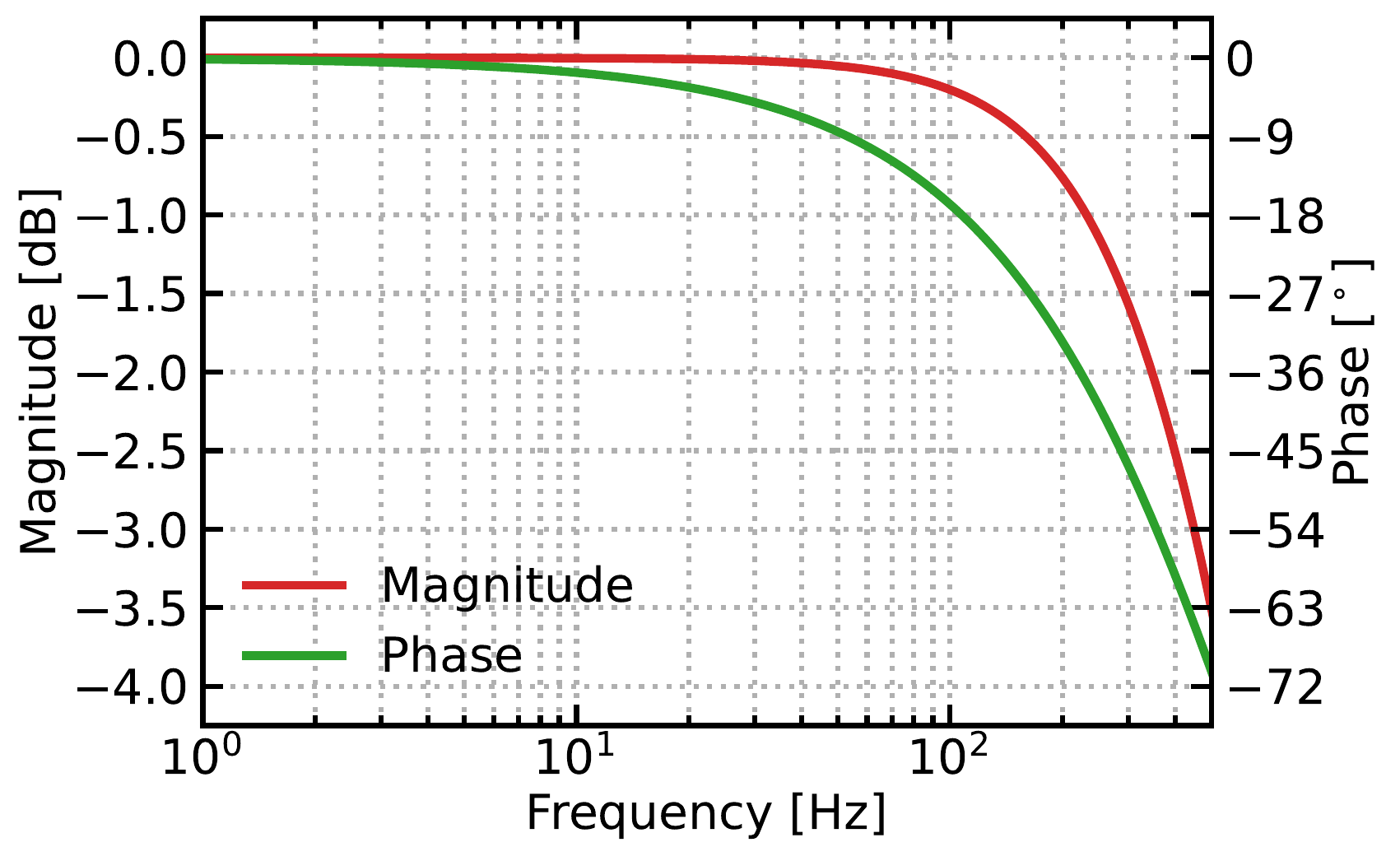}
		\caption{Transfer function of the DAC, showing the magnitude and phase responses. The left y-axis indicates the magnitude in decibels (dB), while the right y-axis indicates the phase in degrees.}
		\label{tf_dac}
	\end{figure}

    With the given parameters, the transfer function of the digital LPF ($F_{\text{LPF}}(z)$) was derived. Similarly, the transfer function of the servo system ($F_{\text{Servo}}(z)$) was computed using \eqref{F_servo}. A delta-sigma DAC, combining a digital modulator and an analog LPF \cite{pavan2017understanding}, was employed in this experiment. Since the bandwidth of the modulator is significantly higher than that of the analog LPF, the transfer function of the DAC is directly determined by the transfer function of the LPF, as shown in Fig.~\ref{tf_dac}.
    
\begin{figure}[!htp]
		\centering
		\includegraphics[width=0.483\textwidth]{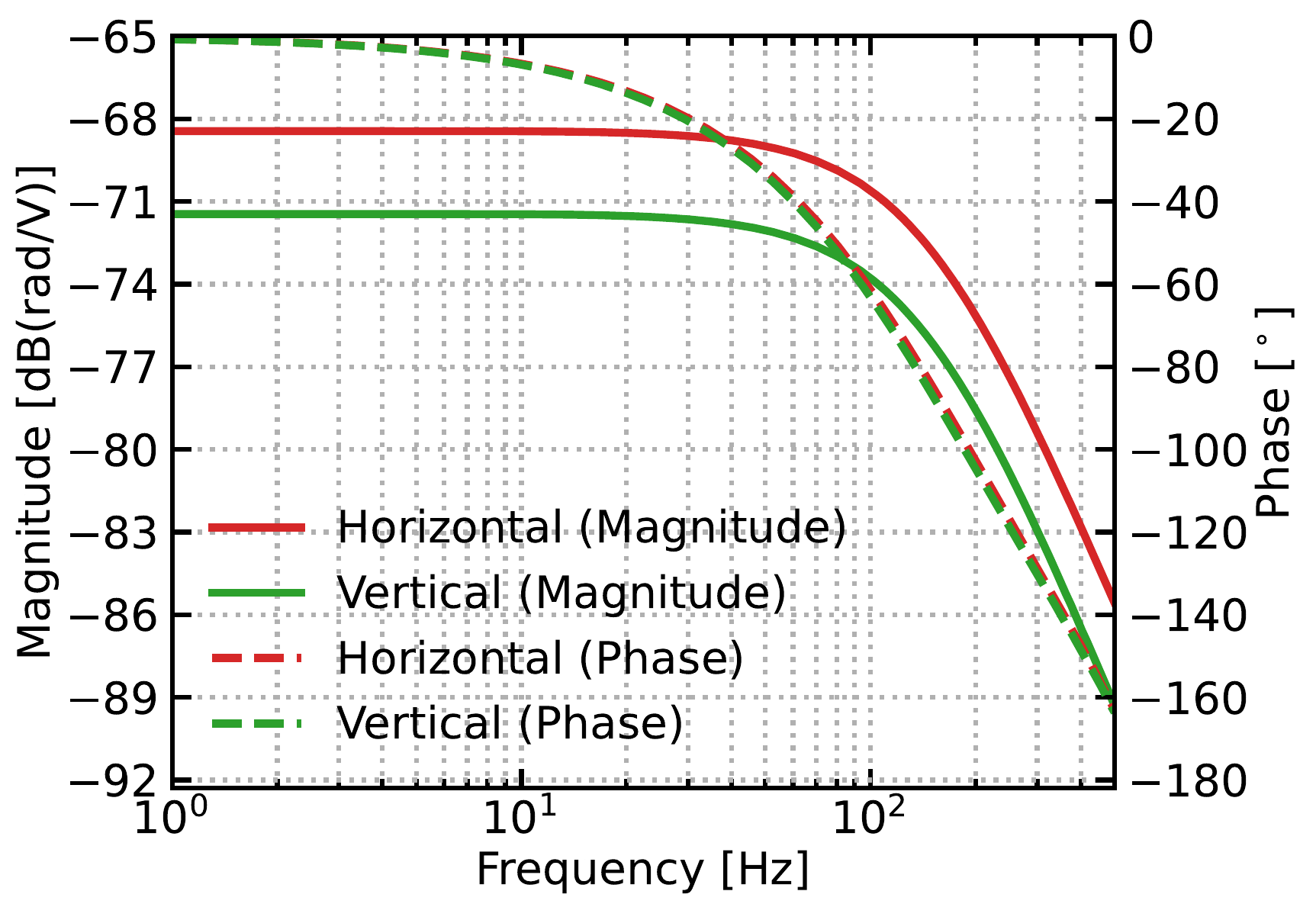}
		\caption{Transfer function of the beam-steering mechanism ($F_{\text{BSM}}$), displaying magnitude and phase responses. The left y-axis indicates the magnitude in dB(rad/V), while the right y-axis indicates the phase in degrees. Contributions from the angular magnification of the imaging system (FSM to QPR), incident angle $\theta_{i}$, steering direction, and FSM frequency response are included. The horizontal and vertical directions correspond to yaw and pitch rotations, respectively.}
		\label{tf_F_BSM}
	\end{figure}

    As explained in \eqref{F_BSM}, the transfer function of the BSM $F_{\text{BSM}}(s)$ is determined by the angular magnification of the imaging system $m$, the sensitivity coefficient $K_{\text{BSM}}$, and the frequency response of the FSM $A(s)$. The angular magnification of the imaging system was derived from its parameters. The nominal incident angle on the FSM is $45^{\circ}$, resulting in $K_{\text{BSM}} = 2$ and $K_{\text{BSM}} = \sqrt{2}$ for the horizontal and vertical steering, respectively, according to \eqref{sensitivity_matrix_nominal}. The horizontal and vertical frequency responses of the FSM were measured using a dynamic signal analyzer (Stanford Research Systems, SR785). Combining the calculated and measured results, the transfer functions of the BSMs were obtained and are plotted in Fig.~\ref{tf_F_BSM}. The horizontal and vertical magnitude responses differ due to their different sensitivity coefficients, whereas the phase responses are nearly identical in both directions.
    
\begin{figure}[!htp]
		\centering
		\includegraphics[width=0.483\textwidth]{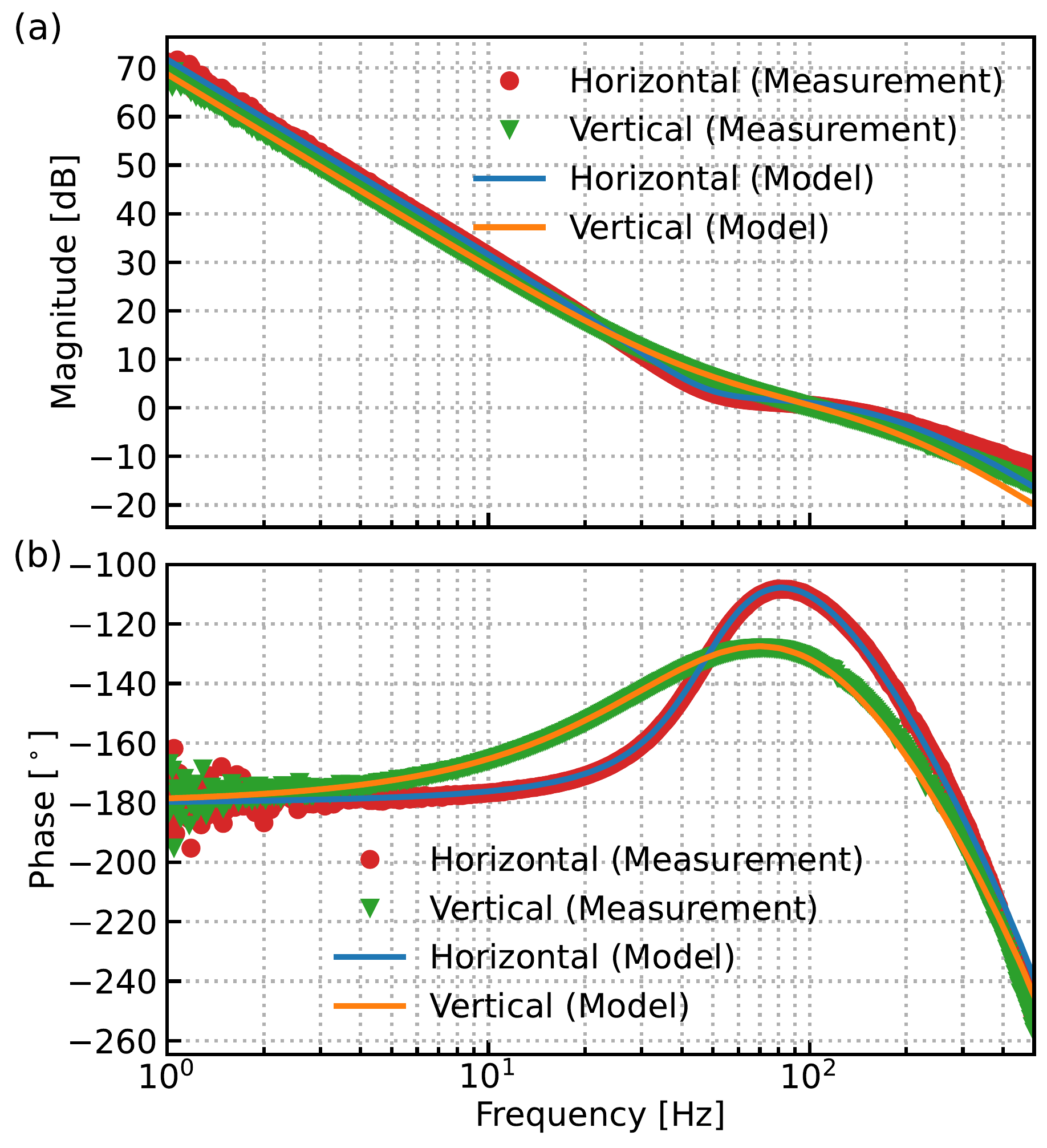}
		\caption{Comparison of measured and modeled open-loop transfer functions, displaying (a) magnitude and (b) phase. The horizontal and vertical results correspond to the BALLs based on the horizontal and vertical DWS signals, respectively. The open-loop transfer functions were measured using a dynamic signal analyzer (Stanford Research Systems, SR785), while the modeled transfer functions were reconstructed using \eqref{G_ol_tf} with the measured or provided transfer functions of the individual components.}
		\label{tf_open_loop}
	\end{figure}

    To characterize the loop transfer functions, frequency-swept sine signals generated by a dynamic signal analyzer were injected at the BSM input (between the $F_{\text{DAC}}$ and $F_{\text{BSM}}$ blocks). Then the corresponding open-loop transfer functions were measured by comparing the DAC output with the BSM input. The measured horizontal and vertical transfer functions are plotted in Fig.~\ref{tf_open_loop}, indicated by the red circles and green triangles, respectively. Accordingly, the closed-loop transfer function and the error transfer function are obtained as well using \eqref{H_cl_tf} and \eqref{E_tf}.
    
    Given the individual component transfer functions, the open-loop transfer functions in both the horizontal and vertical directions were reconstructed using \eqref{G_ol_tf}, as shown in the model results of Fig.~\ref{tf_open_loop}. The model results are in excellent agreement with the experimental data, validating the developed BALL model. As shown in the phase plot of Fig.~\ref{tf_open_loop}, low-frequency deviations are attributed to ambient temperature and air fluctuations during the measurements.
    
    \subsection{Noise and coalignment measurements}\label{noise_coalignment}
    The horizontal and vertical BALLs maintain the coalignment between the RX and TX beams in the transponder-based LRI. Various noise sources couple into the loops, resulting in beam misalignment. Based on the laser interferometric link, this section examines the noise contributions to out-of-loop and in-loop beam-coalignment measurements and compares them with theoretical reconstructions.
    
\begin{figure}[!htp]
	    \centering
		\includegraphics[width=0.483\textwidth]{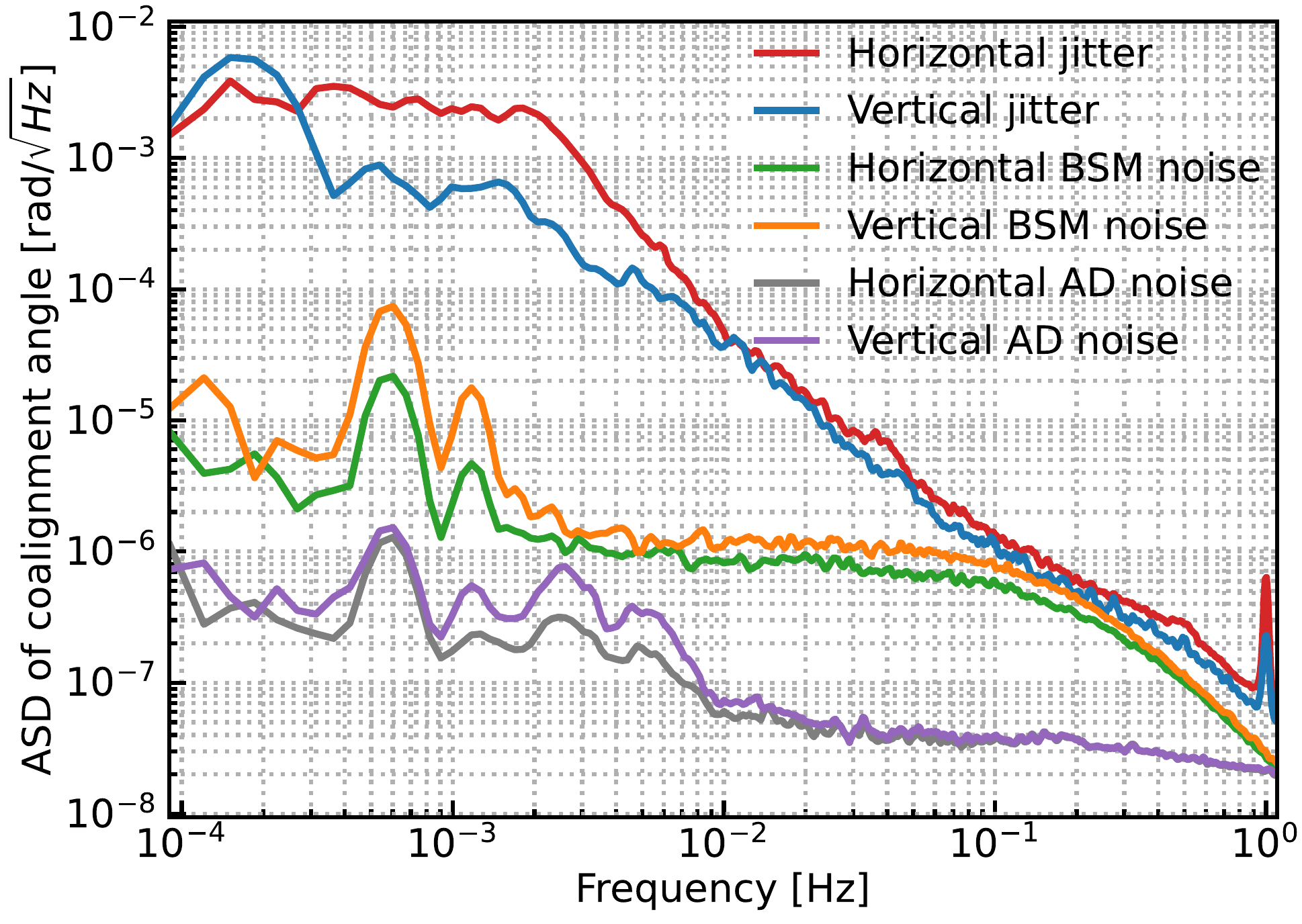}
		\caption{Amplitude spectral density of the measured noise for the horizontal and vertical beam angle-locked loops. The horizontal and vertical jitter correspond to yaw and pitch angular jitter in the reference beams, respectively. These jitters were introduced by hexapod rotations based on GRACE-FO jitter data from January 1, 2019. Unlike the definition in Section \ref{ALL}, the measured BSM noise represents the combined effect of $\widetilde{\theta}_{\text{BSM}}$ and $F_{\text{BSM}}\left(s\right) \cdot \widetilde{U}_{\text{DAC}}$, encompassing contributions from both the FSM and DAC noise. The AD noise shown here corresponds to the measurement of $\widetilde{u}_{\text{h/v}}/F_{\text{AD}}$. AD, angle detector; BSM, beam-steering mechanism.}
		\label{measured_noise}
	\end{figure}

    Angular jitter in the RX beam was produced by hexapod rotations, serving as the reference beam jitter in the BALLs. The GRACE-FO LRI pointing data, recorded on January 1, 2019 \cite{goswami2021analysis}, were used to control the hexapod and simulate spacecraft attitude jitter. The horizontal and vertical loops remained locked during these rotations. The FSM rotation angles were measured via position sensors \cite{FSM_User_Manual}, and the measured angles were converted to RX beam jitter at the aperture using the imaging system's angular magnification. The amplitude spectral densities (ASDs) of the obtained horizontal and vertical jitters are indicated by the red and blue lines, respectively, in Fig.~\ref{measured_noise}.

    A balanced detection scheme (QPR1 and QPR2) was employed in the experiment, as shown in Fig.~\ref{Experimental_setup}. To characterize the BSM noise and the angle detector noise,  a zero-valued digital code was applied to the DAC. With the loops open, the corresponding beam coalignments were measured from the QPR1 and QPR2 DWS signals, which were then converted to misalignment angles using the angle-coupling factors. Based on this configuration, the measured misalignment angles $\widetilde{\delta\theta}_{\text{1}}$ (QPR1) and $\widetilde{\delta\theta}_{\text{2}}$ (QPR2) can be expressed as
    \begin{eqnarray}
        \label{angle_noise_QPR1}
        \widetilde{\delta\theta}_{\text{1}} & = &  \widetilde{\theta}_{\text{1,AD}} +  \widetilde{\theta}_{\text{BSM}}, \\
        \label{angle_noise_QPR2}
        \widetilde{\delta\theta}_{\text{2}} & = &  \widetilde{\theta}_{\text{2,AD}} +  \widetilde{\theta}_{\text{BSM}},
    \end{eqnarray}
    where $\widetilde{\theta}_{\text{1,AD}}$ and $\widetilde{\theta}_{\text{2,AD}}$ represent the equivalent angle noise measured from QPR1 and QPR2, corresponding to the measurement of $\widetilde{u}_{\text{h/v}}/F_{\text{AD}}$ from each QPR; $\widetilde{\theta}_{\text{BSM}}$ represents the equivalent angle noise at the steering beam. The BSM noise $\widetilde{\theta}_{\text{BSM}}$ shown here includes contributions from the DAC and FSM, and is equivalent to the term $\widetilde{\theta}_{\text{BSM}} + F_{\text{BSM}}(s) \widetilde{U}_{\text{DAC}}$ in Section~\ref{ALL}.

    Assuming the noise performance of the angle detector at QPR1 and QPR2 is identical, the ASD of the angle detector noise can be derived from the differential measurement of $\widetilde{\delta\theta}_{\text{1}}$ and $\widetilde{\delta\theta}_{\text{2}}$, which cancels out the common mode noise $\widetilde{\theta}_{\text{BSM}}$. The ASD of the angle detector $\text{ASD}_{\text{AD}} (f)$ is given by
    \begin{equation}
        \label{asd_AD}
        \text{ASD}_{\text{AD}} (f)  =   \sqrt{2} \mathcal{A}\{ \widetilde{\delta\theta}_{\text{1}} - \widetilde{\delta\theta}_{\text{2}} \}, 
    \end{equation}
    where $\mathcal{A}\{\cdot\}$ denotes an operator to calculate the ASD of a time series, which is the square root of the PSD of the given data. Since the BSM noise and the angle detector noise are uncorrelated, the ASD of the BSM noise can be calculated using the relation
    \begin{equation}
        \label{asd_BSM}
        \text{ASD}_{\text{BSM}} (f) = \sqrt{\mathcal{A}^2\{ \frac{\widetilde{\delta\theta}_{\text{1}} + \widetilde{\delta\theta}_{\text{2}}}{2} \} - \text{ASD}^{2}_{\text{AD}} (f)},
    \end{equation}
    with the measured misalignment angles and the ASD of the angle detector noise. Here, the average of $\widetilde{\delta\theta}_{\text{1}}$ and $\widetilde{\delta\theta}_{\text{2}}$ is used to determine the ASD. The determined ASDs of the horizontal and vertical BSM noise and angle detector noise are presented in Fig.~\ref{measured_noise}.

\begin{figure}[!htp]
	    \centering
		\includegraphics[width=0.483\textwidth]{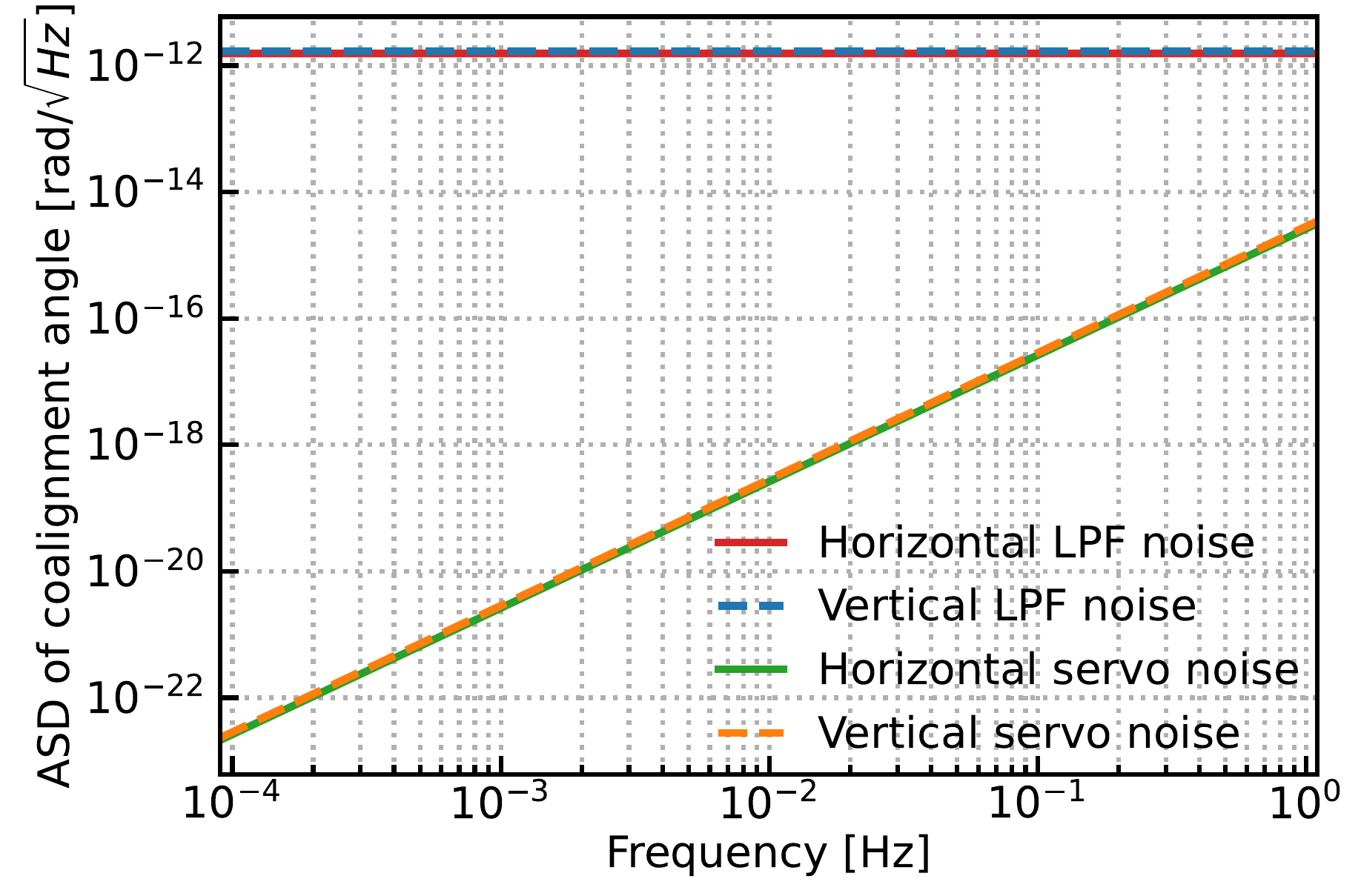}
		\caption{Amplitude spectral density of the estimated low-pass filter (LPF) and servo system noise, including contributions to the horizontal and vertical loops. The noise shown here represents the equivalent sensing noise in the loop. The LPF noise is derived from $\widetilde{u}_{\text{LPF}}/(F_{\text{AD}} \cdot F_{\text{LPF}}(z))$, while the servo noise is derived from $\widetilde{u}_{\text{Servo}}/(F_{\text{AD}} \cdot F_{\text{LPF}}(z) \cdot F_{\text{Servo}}(z))$.}
		\label{servo_lpf_noise}
	\end{figure}

    In addition to the measured noise, digital processing introduces noise through truncation, rounding, and dithering in the LPF and servo system. The ASD of the noise introduced by these operations can be estimated by
    \begin{equation}
       \mathcal{A}\{\widetilde{x}\} =  \frac{2^{-X}\sqrt{3}}{\sqrt{6f_{\text{s}}}},
       \label{asd_trunc_dith}
    \end{equation}
    with the sampling frequency $f_s$ and the digital bit length $X$ \cite{gerberding2014phase, bode2024noise}. The factor of $\sqrt{3}$ accounts for the combined contributions of dithering and truncation noise. The estimated ASDs of the LPF noise and the servo system noise are shown in Fig.~\ref{servo_lpf_noise}. Note that the estimated noise shown here is the sensing noise in the loop, computed via $\widetilde{u}_{\text{LPF}}/(F_{\text{AD}} \cdot F_{\text{LPF}}(z))$ for the LPF noise and $\widetilde{u}_{\text{Servo}}/(F_{\text{AD}} \cdot F_{\text{LPF}}(z) \cdot F_{\text{Servo}}(z))$ for the servo system noise. The LPF noise is white, while the servo system noise increases with frequency. Compared to the measured results in Fig.~\ref{measured_noise}, the contributions from the LPF noise and servo system noise are much less significant.
    
\begin{figure}[!htp]
	    \centering
		\includegraphics[width=0.483\textwidth]{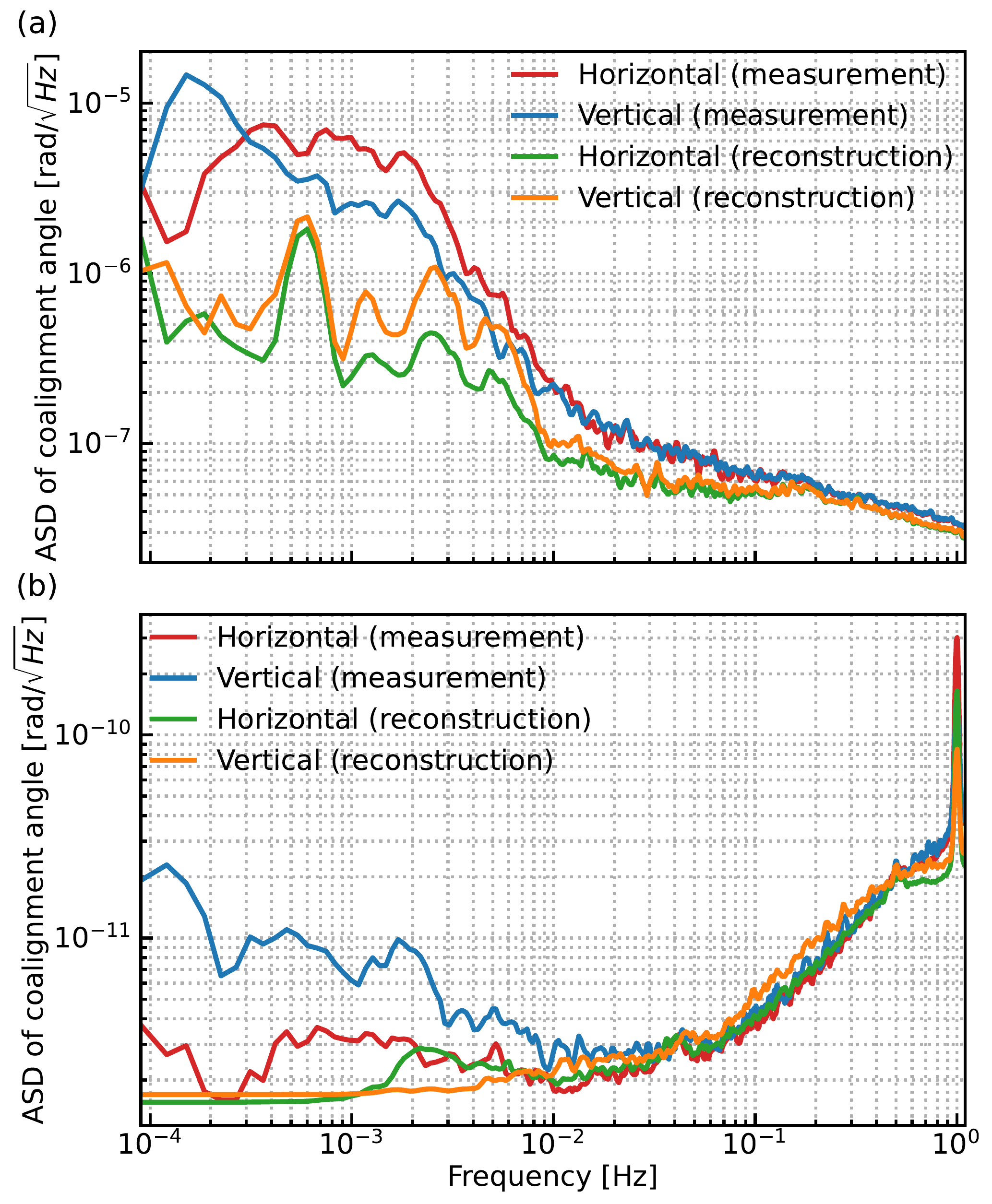}
		\caption{Comparison of measured and reconstructed (a) out-of-loop and (b) in-loop coalignment performance for the horizontal and vertical directions, characterized by the amplitude spectral density. The directly measured data were acquired from QPR2 (out-of-loop) and QPR1 (in-loop) DWS signals and converted to angular deviations using the angle-coupling factor. The theoretical reconstructions were computed using \eqref{PSD_out_loop_error} and \eqref{PSD_in_loop_error}, respectively, incorporating the modeled open-loop transfer function (Fig.~\ref{tf_open_loop}), measured noise (Fig.~\ref{measured_noise}), and estimated noise (Fig.~\ref{servo_lpf_noise}).}
		\label{measured_loop_error}
	\end{figure}

    With the hexapod rotations and the closed BALLs, more than 9 hours of continuous measurements were conducted. The DWS signal measurements from QPR2 and QPR1 provided the out-of-loop and in-loop measurements, respectively. The measured DWS signals were then converted to angle errors between the two beams using their respective angle-coupling factors. The out-of-loop and in-loop angle errors are presented in Fig.~\ref{measured_loop_error} (a) and (b), respectively, as indicated by the red and blue lines for the horizontal and vertical directions. The measured in-loop angle errors are below $10^{-10}$ rad/$\sqrt{\text{Hz}}$ for both loops, indicating that the BALLs effectively suppress the measured beam misalignment. The out-of-loop measurement shows the genuine beam coalignment between the RX and TX beams. Closing both loops suppresses the coalignment jitter, as evidenced by the reduction in the level of the red and blue lines from Fig.~\ref{measured_noise} to Fig.~\ref{measured_loop_error} (a).
    
    Using the transfer functions determined in Fig.~\ref{tf_open_loop}, the measured noise, and the estimated noise, the out-of-loop and in-loop angle errors were reconstructed via \eqref{out_of_loop_error}, \eqref{in_loop_error}, and the square root relation between the PSD and ASD. In this analysis, we assume that $S_{u,\text{h/v}}(f)$ and $S_{u,\text{m}}(f)$ are identical, and both of them were derived from the measured $\text{ASD}_{\text{AD}}(f)$. In comparison to the measured results, the reconstruction results are plotted in Fig.~\ref{measured_loop_error} (a) and (b) as well. The green and orange lines represent the results for the horizontal and vertical directions, respectively.

    As shown in Fig.~\ref{measured_loop_error} (a), the reconstructed out-of-loop angle errors agree well with the measurement results for frequencies higher than 2 mHz. The slight discrepancies in this frequency range mainly result from underestimating the angle detector noise. Since the experimental setup was exposed to air and no temperature control was implemented, the differences between the measured and reconstructed results below 2 mHz are attributed to thermal and air fluctuations. These fluctuations destabilized the interferometric system and introduced noise into the phasemeter, ultimately causing the angle detector transfer function $F_{\text{AD}}$ to vary with ambient conditions rather than remain a constant factor in the reconstruction.
    
    The out-of-loop measurement shows that the coalignment between the RX and TX beams is better than 10 µrad$/\sqrt{\text{Hz}}$ within a frequency range from 0.1 mHz to 1 Hz for both directions, except for a slight excess above 10 µrad$/\sqrt{\text{Hz}}$ between 0.1 mHz and 0.2 mHz for the vertical direction. As explained in \eqref{PSD_out_loop_error}, the contributions from the reference beam jitter and the BSM noise have been significantly suppressed by the BALLs via the error transfer function $E(s, z)$. Compared to the remaining LPF noise and servo system noise, the angle detector noise constrains the coalignment performance of the two beams, as evidenced by the shape similarity in the ASD curves of the angle detector noise (Fig.~\ref{measured_noise}) and the out-of-loop angle error (Fig.~\ref{measured_loop_error}). To further improve beam coalignment stability, the experimental setup must be implemented in a vacuum chamber with robust temperature stabilization, thereby reducing angle detector noise.

    Figure~\ref{measured_loop_error} (b) compares the measured in-loop angle error with its theoretical reconstruction, revealing consistent behavior across both frequency regimes. At high frequencies, the error increases with frequency, whereas at low frequencies, it asymptotically approaches a plateau. Unlike the out-of-loop case, the contribution from the angle detector noise was also significantly suppressed by the loop error transfer function. The low-frequency plateau was dominated by contributions from the LPF and the servo system noise. The reconstructed and measured results exhibit excellent agreement at high frequencies for both steering directions, but discrepancies emerge below 10 mHz. These low-frequency deviations result from ambient temperature and air fluctuations during the experiment. The dependence of the angle detector transfer function $F_{\text{AD}}$ on ambient conditions causes the reconstructed LPF and servo noise contributions to diverge from the experimental data in this regime.

    \subsection{Loop optimization}\label{Sec_loop_optimization}
    The linear transfer function and noise models of the BALL have been established and experimentally validated. Leveraging these models, the loop performance can be optimized by tuning the open-loop transfer function $G(s,z)$ without hardware modifications. In the context of interspacecraft laser interferometry, beam jitter arising from spacecraft attitude fluctuations, angle detector noise, and beam-steering mechanism (BSM) noise can be characterized using onboard measurements. By integrating these measured noise spectra with the theoretical model, the optimal loop gain can be determined to maximize coalignment performance.

    To quantify the beam coalignment over a specified frequency band $[f_0, f_1]$, a standard deviation of the angle error $\sigma_{\text{err}} $ is defined as
    \begin{equation}
      \sigma_{\text{err}} = \sqrt{\int_{f_0}^{f_1} \text{ASD}^2_{\theta}(f)},
       \label{Std_dev_angle}
    \end{equation}
    where $\text{ASD}_{\theta}(f)$ denotes the ASD of the out-of-loop angle error. The optimization objective is to minimize $\sigma_{\text{err}}$ within the target frequency band while ensuring loop stability. Loop stability is assessed via the unity-gain frequency and phase margin of the open-loop transfer function.

    \begin{figure}[!htp]
	    \centering
		\includegraphics[width=0.483\textwidth]{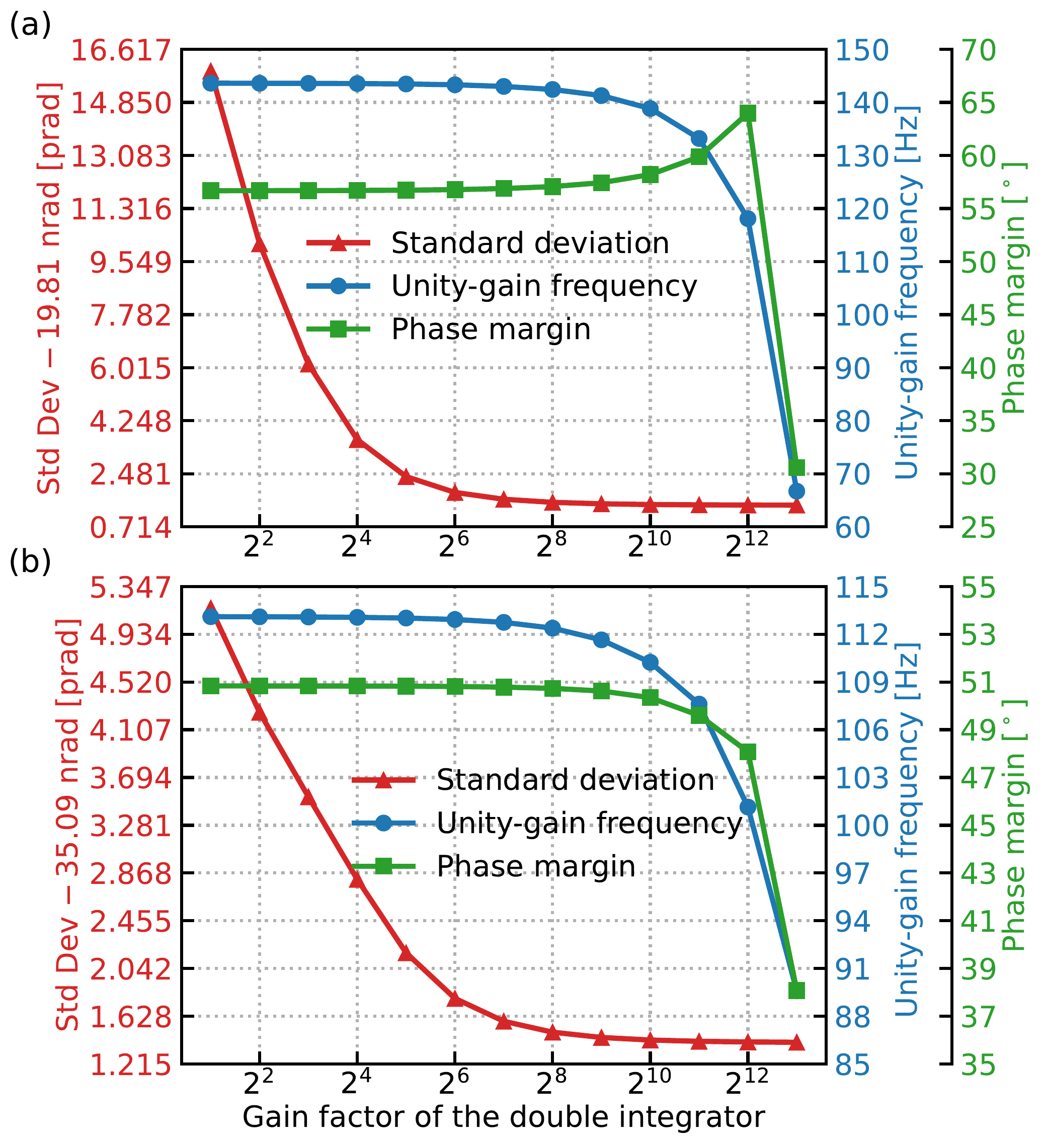}
		\caption{Standard deviation of the angle error, loop unity-gain frequency, and phase margin as functions of the double-integrator gain factor for the (a) horizontal and (b) vertical loops. The left y-axis (red) shows the standard deviation of the misalignment angle in picoradians, the first right y-axis (blue) indicates the loop unity-gain frequency, and the second right y-axis (green) indicates the phase margin. The standard deviation is calculated over the target frequency band from 1 mHz to 100 mHz. To emphasize the small variation with gain, a constant offset of 19.81 nrad (a) and 35.09 nrad (b) has been subtracted from the standard deviation plotted on the left y-axis, as indicated in the corresponding axis label.}
		\label{loop_optimization}
	\end{figure}
    This experiment is a demonstrator for interspacecraft laser interferometry, targeting  frequencies below 100 mHz. Because notable discrepancies arise between the model and measurements at frequencies below 1 mHz, we selected the 1--100 mHz band for loop optimization. The BALL loop gain is effectively tuned by adjusting the double-integrator gain factor $g_{ii}$. Using the measured and estimated noise ASDs alongside the transfer function model, the out-of-loop angle error is computed via \eqref{PSD_out_loop_error}. For various values of $g_{ii}$, the standard deviation is evaluated over the 1--100 mHz band, as shown in Figs.~\ref{loop_optimization}(a) and (b) for the horizontal and vertical loops, respectively. The corresponding unity-gain frequencies and phase margins are also calculated, as indicated by the blue circles and green squares in Fig.~\ref{loop_optimization}.

    For both loops, the angle error standard deviation decreases with increasing gain and approaches a plateau. Conversely, the unity-gain frequency remains relatively insensitive to initial gain increases but decreases sharply when $g_{ii} > 2^{10}$. The phase margins exhibit different behaviors for the horizontal and vertical loops. For the horizontal loop, the margin slightly increases with gain before dropping significantly above $2^{12}$, whereas for the vertical loop, it decreases monotonically as the gain increases. The optimal double-integrator gain factor is determined to be $2^{12}$ for both axes. At this gain, the standard deviation reaches its plateau, and the phase margins (63.99$^{\circ}$ and 48.08$^{\circ}$ for the horizontal and vertical loops, respectively) exceed the typical stability threshold of 45$^{\circ}$. Furthermore, the unity-gain frequencies for both loops are well above the spacecraft attitude jitter frequency, which typically remains below 1 Hz \cite{daniel_intersatellite_2015}.

    In this experiment, the angle detector noise is the dominant contributor to the out-of-loop angle error, coupling through the closed-loop transfer function $H(s,z)$. Consequently, the standard deviation of the angle error decreases monotonically as the open-loop gain increases. If the angle detector noise were further reduced, for instance, by housing the system in a vacuum chamber with improved thermal stabilization, a minimum in the standard deviation would emerge at an optimal gain value. In such a scenario, loop stability constraints would dictate the final selection of the gain. Specifically, the unity-gain frequency must exceed the relevant noise bandwidth, and a sufficient phase margin is required to prevent instability.

    \section{Discussion and conclusion}\label{dis_con}
    This work presents a comprehensive theoretical and experimental framework for the BALL, which actively maintains beam coalignment in laser interferometry. The BALL leverages the DWS signal to extract minute angular misalignments, using a linear relationship with an angle-coupling factor on the order of $10^3$ to $10^4$. These DWS signals are obtained using an angle detector, an interferometer system that measures the phase difference between segments of the QPR. The output of the angle detector provides the error signal for a feedback control chain comprising an LPF, a servo system, a DAC, and a BSM. Through a linear model, we derived rigorous transfer functions and a noise-propagation framework that quantifies the contributions of individual components. This analysis establishes the connection between the PSD of the output beam angle and the PSD of each loop element, specifically detailing the noise contributions to both out-of-loop and in-loop angle errors.

    The model was experimentally validated using an on-axis LRI-like system spanning two optical benches. The reference bench mounted on a hexapod simulated spacecraft attitude jitter, introducing angular jitter into the reference beam and producing misalignments between the RX and TX beams. These misalignments were actively suppressed by two independent BALLs configured for horizontal and vertical steering, which utilize an FSM on the optical bench as the BSM to compensate for beam jitter. The open-loop transfer functions of both loops were characterized, showing excellent agreement with modeling curves. With the closed BALLs and applied attitude jitter, the coalignment between the RX and TX beams was measured, achieving a pointing stability of 10 µrad$/\sqrt{\text{Hz}}$ between 0.2 mHz and 1 Hz. In addition, the noise performance of each loop component, including beam jitter, BSM noise, angle detector noise, LPF noise, and servo noise, was experimentally measured or derived from specified parameters. Using these determined noise spectra and the developed model, the out-of-loop and in-loop angle errors were reconstructed. These predictions align well with the measured coalignment results, with discrepancies below 10 mHz attributed to thermal and air fluctuations not included in the modeling.

    The implementation of the PII controllers enables the BALLs to achieve very high gain, which effectively suppresses noise contributions from reference beam jitter and BSM imperfections within the control bandwidth. Our noise analysis reveals that the out-of-loop coalignment performance is limited by the angle detector noise, which couples through the closed-loop response, whereas the in-loop error at low frequencies is dominated by quantization noise from the digital LPF. To further enhance coalignment performance, it is necessary to suppress angle detector noise, which can be achieved by enclosing the interferometer in a vacuum chamber with stable temperature control.

    Using the analytical model, we demonstrated a loop-optimization strategy that minimizes angle misalignment within a specific frequency range while preserving robust stability margins. This optimization enhanced loop performance by tuning the servo system's gain factors without requiring hardware modifications. The strategy is particularly well-suited for interspacecraft interferometry, where onboard noise measurements reflect the performance of individual loop components. By integrating these measurements with the model, the BALLs can be optimized.

    Although this experimental work is based on a heterodyne interferometer, the BALL concept and technique are applicable to other interferometric architectures, such as homodyne, deep phase modulation (DPM), and deep frequency modulation (DFM) interferometers. For these architectures, the primary distinction lies in how the angle detector is implemented. Similar control loops have been successfully implemented in ground-based homodyne interferometers \cite{fritschel1998alignment,heinzel_automatic_1999,grote2004alignment}, and DPM and DFM interferometers are also capable of generating DWS signals \cite{heinzel2010deep,isleif2019compact}. Furthermore, the BALL concept can be extended to single-beam pointing stabilization by replacing DWS signals with differential power sensing signals or by utilizing an array detector to extract angular information. This extension is particularly relevant for pointing control in interspacecraft and space-to-ground laser communication. In this configuration, the beam is aligned to the center of the QPR or detector, rather than relative to a reference beam as in the present work.
    
    The validated BALL model offers a framework for designing and optimizing active beam alignment in laser interferometry. However, certain limitations still exist. The model relies on the assumption of linear frequency responses for all loop components, and low-frequency angle error discrepancies were observed due to thermal and air fluctuations. Future work should focus on conducting experiments in vacuum environments with active thermal stabilization, thereby facilitating more precise comparisons at low frequencies. Additionally, the potential nonlinear effects and behaviors of the loop component should be analyzed.

    \appendix
    \titleformat{\section}
    {\large\sffamily\bfseries}
    {APPENDIX \thesection:}
    {0.5em}
    {\MakeUppercase{#1}}

    \counterwithin{equation}{section}
    \renewcommand{\theequation}{\thesection\arabic{equation}}

 	\section{Angle detection}\label{app_angle_detection}
	To describe the DWS angle-coupling factor for the case of a finite QPR, two functions of $\eta$ are defined below
    \begin{equation}
    \label{F_0_beta}
    F_{0}\left(\eta \right)  =  \frac{\sqrt{2}\left( \sqrt{\pi} \mathrm{erf}\left( \eta \right)-2\eta e^{-\eta^{2}}\right)}{\pi\left( 1-e^{-\eta^{2}} \right)}, \\
    \end{equation}
    \begin{eqnarray}
    &F_{2}\left(\eta \right)&  =  \frac{{\sum\limits_{n=0}^{3} e^{-n\eta^{2}}}\iota_n}{2^{\frac{3}{2}}\pi\left( 1-e^{-\eta^{2}} \right)},  \nonumber\\
    \iota_0 &=& -3\sqrt{\pi} \text{erf}\left( \eta\right), \nonumber\\
    \iota_1 &=& 6\eta+4\eta^3+8\eta^5 +\sqrt{\pi}\text{erf}\left( \eta\right)\left( 6-4\eta^2-4\eta^4\right), \nonumber\\
    \iota_2 &=&-12\eta+8\eta^5 +\sqrt{\pi}\text{erf}\left( \eta\right)\left( -3+4\eta^2-4\eta^4\right), \nonumber \\
    \iota_3 &=& 6\eta-4\eta^3,
    \label{F_2_beta}
    \end{eqnarray}
    where $\eta$, defined in \eqref{beta_qpr}, characterizes the relative magnitude of the QPR active area to the effective beam size on the QPR. The function $\text{erf}\left( \eta\right)$ denotes the error function of $\eta$, which is defined as
    \begin{equation}
    \label{error_function}
    \text{erf} \left ( \eta \right )  = \frac{2}{\sqrt{\pi} }\int_{0}^{\eta} e^{-t^2} dt. 
    \end{equation}
    Note that the results presented here are derived from Ref.~\cite{pizzella2025mathematical}.

	\section{DWS correction matrix}\label{app_DWS matrix}
    Misalignments of the QPR result in discrepancies between the definitions of the horizontal and vertical axes in the QPR frame and the beam steering frame. Consequently, cross-coupling appears in the measured DWS signals. To eliminate this effect, an experimentally determined matrix is employed. This section discusses the derivation and application of this correction matrix.  
    
    \begin{figure}[!htp]
		\centering
		\includegraphics[width=0.45\textwidth]{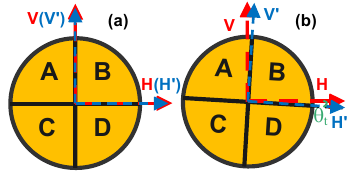}
		\caption{Sketch explaining QPR alignment. In the aligned case (a), two coordinate frames coincide. In the misaligned case (b), the QPR is tilted by $\theta_{t}$. The red axes (H, V) represent horizontal and vertical directions in the beam steering frame, while the blue axes ($\text{H}^{'}$, $\text{V}^{'}$), fixed to the QPR, define the directions of the measured DWS signals.}
		\label{mis_qpr}
	\end{figure}
    As illustrated in Fig.~\ref{mis_qpr}(a), the horizontal and vertical axes are identical in both the beam steering frame and the QPR frame. However, QPR misalignments cause an angular mismatch of $\theta_{t}$, as shown in Fig.~\ref{mis_qpr}(b). Under this misalignment, the measured horizontal and vertical DWS signals, denoted $\mathrm{DWS_{h}^m}$ and $\mathrm{DWS_{v}^m}$, are given by
    \begin{equation}
    \begin{aligned}
    \begin{pmatrix} \mathrm{DWS_{h}^m} \\ \mathrm{DWS_{v}^m} \end{pmatrix}
    &= \begin{pmatrix} \kappa_h & 0 \\ 0 & \kappa_v \end{pmatrix} 
    \begin{pmatrix} \cos\theta_{t} & -\sin\theta_{t} \\ \sin\theta_{t} & \cos\theta_{t} \end{pmatrix}
    \begin{pmatrix} \alpha \\ \beta \end{pmatrix}\\
    &= \begin{pmatrix} \kappa_h\cos\theta_{t} & -\kappa_h\sin\theta_{t} \\ \kappa_v\sin\theta_{t} & \kappa_v\cos\theta_{t} \end{pmatrix}\begin{pmatrix}  \alpha \\  \beta \end{pmatrix},
    \end{aligned}
    \label{DWS_measurements}
    \end{equation}
    where $\kappa_h$ and $\kappa_v$ are the horizontal and vertical DWS angle-coupling factors, respectively, with the axes defined in the QPR frame. The quantities $ \alpha$ and $ \beta$ represent the yaw and pitch misalignment angles between two laser beams. The central matrix of the right side of the equation represents a rotation by $\theta_t$, which transforms the beam misalignment angles from the beam steering frame to the QPR frame. This rotation introduces cross-coupling between the two orthogonal directions, characterized by the off-diagonal terms $-\kappa_h \sin\theta_{t}$ and $\kappa_v \sin\theta_{t}$, which is undesirable for the two independent BALLs.
    
    To eliminate this cross-coupling, we define a correction matrix 
    \begin{equation}
        \mathrm{M_{c}} = \begin{pmatrix}\cos\theta_{t} & \frac{\kappa_h}{\kappa_v}\sin\theta_{t} \\ -\frac{\kappa_v}{\kappa_h}\sin\theta_{t} & \kappa_v\cos\theta_{t} \end{pmatrix}.
        \label{correction_matrix}
    \end{equation}
    Multiplying \eqref{DWS_measurements} by $\mathrm{M_{c}}$ yields
    \begin{equation}
    \begin{pmatrix} \mathrm{DWS_{h}^f} \\ \mathrm{DWS_{v}^f} \end{pmatrix}
    = \mathrm{M_{c}}
    \begin{pmatrix} \mathrm{DWS_{h}^m} \\ \mathrm{DWS_{v}^m} \end{pmatrix}
    = \begin{pmatrix} \kappa_h & 0 \\ 0 & \kappa_v \end{pmatrix}\begin{pmatrix}  \alpha \\  \beta \end{pmatrix},
    \label{DWS_feedback}
    \end{equation}    
    where $\mathrm{DWS_{h}^f}$ and $\mathrm{DWS_{v}^f}$ represent the decoupled DWS signals. These signals provide feedback to the BALLs, which correct horizontal and vertical beam misalignments. 

    To determine the correction matrix, beam misalignments are manually introduced using the BSM while the DWS signals are simultaneously recorded. By fitting the measured DWS signals to the misalignment angles and utilizing the relationship in \eqref{DWS_measurements}, the angle-coupling factors ($\kappa_h$ and $\kappa_h$) and the QPR misalignment angle ($\theta_t$) can be determined. These parameters are then used to construct the correction matrix $\mathrm{M_{c}}$ via \eqref{correction_matrix}.

    \section{Beam-steering mechanism of FSM}\label{app_fsm_bsm}
    In this section, we derive a model describing the relationship between the FSM rotation angle and the angular deflection of the laser beam. The derived sensitivity matrix relates the two angles and is used in Section \ref{ALL} to describe the transfer function of the BSM.

    \begin{figure}[!htb]
    \centering
    \includegraphics[width=0.45\textwidth]{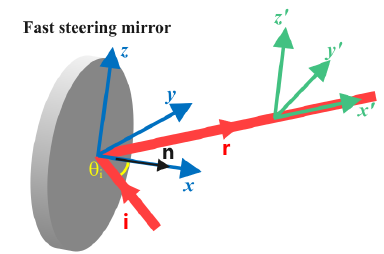}
    \caption{FSM frame $\mathbf{xyz}$ (blue) and reflected‑beam frame $\mathbf{x'y'z'}$ (gree). The $\mathbf{z'}$ axis is aligned with the mirror $\mathbf{z}$ axis, while the $\mathbf{y'}$ axis lies in the $\mathbf{xy}$ plane. The unit vectors
    $\mathbf{i}$, $\mathbf{r}$, and $\mathbf{n}$ denote the incident beam, reflective beam, and mirror surface normal. The incident angle of the beam at the mirror is $\theta_{i}$.}
    \label{coordinate_fsm_bsm}
    \end{figure}
    
    As shown in Fig.~\ref{coordinate_fsm_bsm}, a laser beam reflects off an FSM with an incident angle of $\theta_{in}$. To model the reflection process, we define two coordinate frames: the FSM frame $xyz$ and the reflected-beam frame $x'y'z'$. The FSM surface lies in the $yz$-plane, such that the surface normal $\mathbf{n}$ points along the $x$-axis. The nominal incident beam propagates in the $xy$-plane at an angle $\theta_i$ relative to the normal, given by
    \begin{equation}
    \mathbf{i}_0 = (-\cos\theta_i,\; \sin\theta_i,\; 0),
    \end{equation}
    while the nominal reflected beam is
    \begin{equation}
    \mathbf{r}_0 = (\cos\theta_i,\; \sin\theta_i,\; 0).
    \label{vec_r_ori}
    \end{equation}

    The beam-fixed frame $x'y'z'$ is attached to the reflected beam, with its longitudinal axis aligned with $\mathbf{r}_0$. The $z'$-axis is parallel to the FSM's $z$-axis, yielding
    \begin{equation}
    \mathbf{z}' = (0,\; 0,\; 1), \qquad
    \mathbf{y}' = \mathbf{z}' \times \mathbf{x}' = (-\sin\theta_i,\; \cos\theta_i,\; 0).
    \label{axis_beam_frame}
    \end{equation}
    For small angular deviations, the change in the reflected beam direction, $\Delta\mathbf{r}$, is approximated by
    \begin{equation}
    \Delta\mathbf{r} \approx \boldsymbol{\theta}_r \times \mathbf{r},
    \label{small_angle_change}
    \end{equation}
    where $\boldsymbol{\theta}_r = (\gamma, \beta, \alpha)$ is the rotation vector expressed in the $x'y'z'$ frame. The components $\gamma$, $\beta$, and $\alpha$ represent the small rotation angles abou $x'$, $y'$, and $z'$ axes, respectively. Substituting $\mathbf{r}_0 = \mathbf{x}'$ into \eqref{small_angle_change} and projecting onto the transverse axes yields the yaw and pitch angles:
    \begin{equation}
    \alpha \approx \Delta\mathbf{r} \cdot \mathbf{y}', \qquad
    \beta \approx -\Delta\mathbf{r} \cdot \mathbf{z}'.
    \label{eq:proj}
    \end{equation}    

    When the FSM undergoes small rotations $\theta_{\text{yaw}}$ about its $z$-axis and $\theta_{\text{pitch}}$ about its $y$-axis, the surface normal to first order becomes
    \begin{equation}
    \mathbf{n} \approx (1,\; \theta_{\text{yaw}},\; -\theta_{\text{pitch}}).
    \end{equation}
    Applying the law of reflection, $\mathbf{r} = \mathbf{i} - 2(\mathbf{i} \cdot \mathbf{n})\mathbf{n}$, with the nominal incident direction $\mathbf{i}_0$ yields, to first order,
    \begin{equation}
    \begin{aligned}
    \mathbf{r} \approx& \bigl( \cos\theta_i - 2\theta_{\text{yaw}}\sin\theta_i,\; \sin\theta_i + 2\theta_{\text{yaw}}\cos\theta_i,\; \\
    &-2\theta_{\text{pitch}}\cos\theta_i \bigr).
    \end{aligned}
    \label{r_vector}
    \end{equation}
    Subtracting \eqref{vec_r_ori} from \eqref{r_vector} gives $\Delta\mathbf{r} = (-2\theta_{\text{yaw}}\sin\theta_i,\; 2\theta_{\text{yaw}}\cos\theta_i,\; -2\theta_{\text{pitch}}\cos\theta_i)$. Combining this with \eqref{axis_beam_frame} and \eqref{eq:proj}, the beam deflection angles are found to be
    \begin{equation}
    \alpha = 2\theta_{\mathrm{yaw}}, \qquad
    \beta = 2\cos\theta_i\theta_{\mathrm{pitch}}.
    \label{eq:mirror}
    \end{equation}
    Thus, the sensitivity matrix relating the FSM rotation angles to the beam deflection angles is
    \begin{equation}
    \begin{pmatrix} \alpha \\ \beta \end{pmatrix}
    = \begin{pmatrix} 2 & 0 \\ 0 & 2\cos\theta_i \end{pmatrix}
    \begin{pmatrix} \theta_{\mathrm{yaw}} \\ \theta_{\mathrm{pitch}} \end{pmatrix}.
    \label{sensitivity_matrix_nominal}
    \end{equation}

    We now consider a more general case that includes small incident beam misalignments, characterized by rotations $\theta_u$ about the FSM $x$-axis and $\theta_v$ about the $y$-axis. To first order, the incident direction becomes
    \begin{equation}
    \mathbf{i} \approx \mathbf{i}_0 + (0,\; 0,\; \theta_u\sin\theta_i + \theta_v\cos\theta_i).
    \end{equation}
    Re-evaluating the reflection law with this misaligned incident beam and the tilted normal yields the linearized reflected beam:
    \begin{equation}
    \begin{aligned}
    \mathbf{r} \approx \bigl(\cos\theta_i - 2\theta_{\mathrm{yaw}}\sin\theta_i,\;
    \sin\theta_i + 2\theta_{\mathrm{yaw}}\cos\theta_i,\; \\
     \theta_u\sin\theta_i + \theta_v\cos\theta_i - 2\theta_{\mathrm{pitch}}\cos\theta_i \bigr).
     \end{aligned}
    \end{equation}
    The deviation from the nominal direction is therefore
    \begin{equation}
    \begin{aligned}
    \Delta\mathbf{r} &= \bigl( -2\theta_{\mathrm{yaw}}\sin\theta_i,\; 2\theta_{\mathrm{yaw}}\cos\theta_i,\; \\
    &\theta_u\sin\theta_i + \theta_v\cos\theta_i - 2\theta_{\mathrm{pitch}}\cos\theta_i \bigr).
    \end{aligned}
    \end{equation}
    Applying the projection relations in \eqref{eq:proj} yields the general linearized model:
    \begin{align}
    \alpha &= 2\theta_{\mathrm{yaw}}, \label{eq:yaw_gen} \\
    \beta &= 2\cos\theta_i\theta_{\mathrm{pitch}} - \theta_u\sin\theta_i - \theta_v\cos\theta_i. \label{eq:pitch_gen}
    \end{align}
    \eqref{eq:yaw_gen} and \eqref{eq:pitch_gen} reveal that the reflected beam's yaw angle is controlled exclusively by the FSM yaw rotation (with sensitivity of 2), independent of pitch or incident misalignments. In contrast, the beam pitch is driven by the FSM pitch rotation (with sensitivity of $2\cos\theta_i$) and includes a static offset term arising from the incident misalignments $u$ and $v$.

    
\begin{backmatter}
\bmsection{Funding} Max Planck Society (MPG) (M.IF.A.~QOP18098), the Deutsche Forschungsgemeinschaft (DFG) (434617780-SFB 1464), the Relativistic Geodesy 1128, the Clusters of Excellence (EXC2123 No.~390837967, EXC2122 No.~390833453), and the German Aerospace Center (DLR) (No.~50OQ2301).

\bmsection{Acknowledgment} The authors thank Sergio Lozano Althammer and Jos\'e Alberto Ogalde Ortiz for their helpful discussions.

\bmsection{Disclosures} The authors declare no conflicts of interest.


\bmsection{Data availability} Data underlying the results presented in this paper are not publicly available at this time but may be obtained from the authors upon reasonable request.



\end{backmatter}

\bibliography{sample}

\bibliographyfullrefs{sample}

\end{document}